\documentclass[journal]{IEEEtran}
\usepackage{cite}
\usepackage{amsmath}
\usepackage{amssymb}
\usepackage{xcolor}
\usepackage{colortbl}	
\usepackage[ruled,vlined,linesnumbered]{algorithm2e}
\usepackage{comment}
\usepackage{url}

\newcommand{\cA}{{\cal A}}

\newcommand{\cF}{{\cal F}}

\newcommand{\cI}{{\cal I}}

\newcommand{\cM}{{\cal M}}

\newcommand{\cR}{{\cal R}}
\newcommand{\cS}{{\cal S}} 
\newcommand{\cT}{{\cal T}}

\newcommand{\cX}{{\cal X}}

\DeclareMathAlphabet{\mathbfsl}{OT1}{ppl}{b}{it} 

\newcommand{\bfb}{\mathbfsl{b}} 
\newcommand{\bfc}{\mathbfsl{c}}

\newcommand{\bfu}{\mathbfsl{u}} 
\newcommand{\bfv}{\mathbfsl{v}} 
 
\newcommand{\bfx}{\mathbfsl{x}}

\newcommand{\mbC}{{\mathbb C}}

\newcommand{\mbF}{{\mathbb F}}

\newtheorem{definition}{Definition}
\newtheorem{theorem}{Theorem}
\newtheorem{conjecture}{Conjecture}

\newtheorem{proposition}{Proposition}
\newtheorem{lemma}{Lemma}
\newtheorem{example}{Example}

\newcommand{\Tref}[1]{The\-o\-rem\,\ref{#1}}
\newcommand{\Pref}[1]{Pro\-po\-si\-tion\,\ref{#1}}
\newcommand{\Lref}[1]{Lem\-ma\,\ref{#1}}
\newcommand{\Cref}[1]{Co\-ro\-lla\-ry\,\ref{#1}}
\newcommand{\Fref}[1]{Fi\-g\-u\-re\,\ref{#1}}

\definecolor{Gray}{gray}{0.9}
\definecolor{LightCyan}{rgb}{0.88,1,1}

\usepackage{tikz}
\usepackage{pgfplots}
\usetikzlibrary{decorations.pathreplacing}

\renewcommand{\preceq}{\preccurlyeq}
\renewcommand{\succeq}{\succcurlyeq}
\newcommand{\rowind}[1]{[\![#1]\!]}
\newcommand{\deff}{\triangleq}

\newcommand{\Arikan}{Ar{\i}\-kan}

\DeclareMathOperator{\MF}{MF}

\DeclareMathOperator{\ev}{ev}
\DeclareMathOperator{\even}{even}
\DeclareMathOperator{\odd}{odd}
\DeclareMathOperator{\vspan}{span}
\DeclareMathOperator{\LTA}{LTA}
\DeclareMathOperator{\ind}{ind}
\DeclareMathOperator{\ovd}{\prec_{1}} 

\pgfplotsset{compat=1.17}
\begin{document}
%
\title{A Deterministic Algorithm for Computing \\
the Weight Distribution of Polar Code}

\author{Hanwen~Yao,~\IEEEmembership{Member,~IEEE,}
        Arman~Fazeli,~\IEEEmembership{Member,~IEEE,}
        and~Alexander~Vardy,~\IEEEmembership{Fellow,~IEEE}
\thanks{The work of Hanwen Yao, 
Arman Fazeli and Alexander Vardy was supported in part 
by the United States National Science Foundation (NSF) under 
Grants CCF-1405119 and Grant CCF-1719139.
An earlier version of this paper was presented in
part at the 2021 International Symposium on Information Theory (ISIT).
(Corresponding author: Hanwen Yao.)}
\thanks{Hanwen Yao is with Duke University, Durham, NC 27708 USA 
(e-mail: hanwen.yao@duke.edu).}
\thanks{Arman Fazeli is with Apple Inc., Los Altos, CA 94022 USA 
(e-mail: afazelic@ucsd.edu).}
\thanks{Alexander Vardy is with the University of California at San Diego,
San Diego, CA 92093 USA (e-mail: vardy@ece.ucsd.edu).}
}

%

\maketitle

\begin{abstract}
In this work, we present a {deterministic} algorithm for computing
the {entire} weight distribution of polar codes. 
As the first step, we derive an efficient recursive procedure 
to compute the weight distribution that arises in 
successive cancellation decoding of polar codes 
{along any decoding path}. 
This solves the open problem recently posed by
Polyanskaya, Davletshin, and Polyanskii.
Using this recursive procedure, at code length $n$, 
we can compute the weight distribution of 
any {\it polar cosets} in time $O(n^2)$.
We show that any polar code can be represented 
as a disjoint union of such polar cosets;
moreover, this representation extends to 
polar codes with dynamically frozen bits. 
However, the number of polar cosets in such representation
{scales exponentially} with a parameter 
introduced herein, which we call the {\it mixing factor}.
To upper bound the complexity of our algorithm 
for polar codes being decreasing monomial codes, 
we study the range of their mixing factors.
We prove that among all decreasing monomial codes 
with rates at most 1/2, 
self-dual Reed-Muller codes have the largest mixing factors. 
To further reduce the complexity of our algorithm,
we make use of the fact that, as decreasing monomial codes,
polar codes have a large automorphism group. 
That automorphism group includes the block 
lower-triangular affine group (BLTA), 
which in turn contains the lower-triangular affine group (LTA).
We prove that a subgroup of LTA
acts transitively on certain subsets of 
decreasing monomial codes, 
thereby drastically reducing the number of polar cosets 
that we need to evaluate. This complexity reduction
makes it possible to compute the weight distribution of 
polar codes at length $n=128$.
\end{abstract}


\section{Introduction}
\label{sec:intro}
\IEEEPARstart{T}{he} 
weight distribution of an error correction code counts 
the number of codewords in this code of all weights.
The weight distribution is one of the main characteristic 
of a code, useful for analysing its performance under 
maximum-likelihood decoding, and various other decoding algorithms. 
However, computing the weight distribution of a general linear 
code is known to be NP-hard \cite{berlekamp1978inherent}.
Hence, there are very few families of codes 
whose weight distribution is known. 
Some families of codes with known weight distributions 
are Hamming codes, Golay codes, and Reed-Solomon codes. 
For primitive Bose-Chaudhuri-Hocquenghem (BCH) codes 
and the extended primitive BCH codes, 
their weight distributions are known for lengths up to 128 
\cite{desaki1997weight}.
Besides, the weight distributions for primitive BCH codes 
of length 255 and extended primitive BCH codes of length 256
have been computed for code dimension $k\le 71$ and $k\ge 187$
\cite{fujiwara1997weight,fujiwara1993weight,fujiwara2021weight}.
For Reed-Muller codes, their weight distributions are known up 
to length 512, except for the (512,256)
Reed-Muller code \cite{sugino1971weight, sugita1996weight}.
Polar codes, introduced by \Arikan~\cite{arikan2009channel}, 
form the first explicit family of codes that provably 
achieve capacity with efficient encoding and decoding for a 
wide range of channels. 
With the brute-force search and 
the MacWilliams identity \cite{MacWilliams}, 
the weight distribution of polar codes 
can be computed up to length 64. 
For example, the weight distribution of
a (64,32) polar code is computed by brute force 
in \cite{xu2017distance} for code design purpose.
However, the weight distribution of polar codes 
at length 128 is currently not known.

\subsection{Related Prior Work}
Introduced by \Arikan \cite{arikan2009channel}, 
polar code is a binary linear code generated 
by a subset of rows in the polar transformation 
matrix, whose corresponding bit channels have
the smallest Bhattacharyya parameters. 
To the best of our knowledge, 
there are no prior results on how to efficiently 
compute the entire weight distribution of 
polar codes. For crude estimations, 
there are probabilistic methods discussed in 
\cite{valipour2013probabilistic} and 
\cite{zhang2017enhanced}. 
Although we don't know the weight distribution of polar codes, 
we do know their minimal weight, 
and the number of codewords of that weight. 
In the work by Bardet, Dragoi, Otmani and Tillich 
\cite{bardet2016algebraic}, 
they look into a partial order relation 
for the bit channels, and introduce a broader class of codes 
following the partial order called decreasing monomial codes. 
This class of decreasing monomial codes includes polar codes.
They also study the automorphism group of 
decreasing monomial codes from a polynomial formalism, 
and provide an explicit formula for the number of codewords 
of minimal weight. Besides that result, 
there are also ways to estimate the first few 
numbers in the weight distribution of polar codes. 
In the work by Li, Shen and Tse \cite{li2012adaptive}, 
they devise an experiment that evaluates 
the number of low-weight polar codewords. 
In this experiment, an all-zero codeword is transmitted 
in the extremely high SNR regime, 
and the channel output is decoded 
by a list decoder \cite{tal2015list}. 
With a large enough list size, 
the first few non-zero numbers in the weight distribution 
can be estimated by counting the low-weight codewords 
obtained in the list. Later in \cite{liu2014distance}, 
this experiment is improved for a memory constraint computer.
But still, with this approach, only the first few numbers 
in the entire weight distribution can be estimated. 
This approach is also {\it non-exact} in the sense that, 
for a given weight, the number of codewords obtained in the list
only serves as a lower bound 
for the actual number in the weight distribution.

\subsection{Our Contributions}
In this paper, we present a deterministic algorithm that 
computes the {\it exact} weight distribution of polar codes.
We first propose an efficient recursive procedure to 
compute the weight enumerating function of {\it polar cosets} 
to be defined later. 
Those polar cosets arise during the successive cancellation (SC)
decoding process, and their weight distribution can be used to estimate the 
error probabilities of the bit channels.
In two separate works by Niu, Li, and Wu \cite{niu2019polar},
and by Polyanskaya, Davletshin, and Polyanskii \cite{polyanskaya2020weight}, 
algorithms that compute the weight distribution of these polar cosets 
along the all-zero decoding path are proposed. 
However, how to efficiently compute the weight distribution of polar cosets 
along an arbitrary decoding path remains an open problem.
In this work, we solve this problem by establishing a recursive relation 
followed by the weight enumerating functions of those polar cosets.  
Using this recursive relation, 
we can compute the weight distribution of polar cosets 
along arbitrary decoding path in time $O(n^2)$. 

Next, we show that we can represent any polar code 
as a disjoint union of certain polar cosets. 
In this way, we can obtain the weight distribution of the entire code 
by summing up the weight enumerating functions 
of those polar cosets. This representation also extends to 
polar codes with dynamically frozen bits, 
which are first introduced in \cite{trifonov2013dynamic}.
Since any binary linear codes can be represented as polar codes 
with dynamically frozen bits \cite{trifonov2013dynamic}, 
our algorithm applies to general linear codes as well. 
However, the number of polar cosets in this representation scales 
exponentially with a code parameter introduced herein, 
which we call the {mixing factor}.
The complexity of our algorithm is largely governed 
by the mixing factor of polar codes. 

Representing polar codes as disjoint unions of polar cosets 
works for polar codes in a general setting, 
where we can select any subsets of 
rows in the polar transformation matrix as generators.
In a more restricted setting, where we only select the rows 
whose corresponding bit channels have  
the smallest Bhattacharyya parameters, 
polar codes fall into the category of 
decreasing monomial codes \cite{bardet2016algebraic}. 
To upper bound the mixing factor of 
polar codes being decreasing monomial codes, 
and thus give a bound on the complexity of our algorithm, 
we prove that self-dual Reed-Muller codes 
have the largest mixing factor among all 
decreasing monomial codes with rates at most 1/2.

As decreasing monomial codes, polar codes have a large
automorphism group. It is first shown in \cite{bardet2016algebraic} 
that the automophism group of decreasing monomial codes
includes the lower triangular affine group (LTA).
Recently in \cite{geiselhart2021automorphism}, 
this result has been extended to  
the block lower triangular affine group (BLTA).
Later in \cite{li2021complete}, 
it has also been shown that BLTA equals the complete 
automorphisms of decreasing monomial 
codes that can be formulated as affine transformations.
In our work, we show that using 
a subgroup of LTA, we can largely reduce the complexity of our algorithm. 
We prove that the subgroup we considered 
acts transitively on certain subsets of decreasing monomial codes, 
which implies that a lot of polar cosets in our 
representation share the same weight distribution. 
It allows us to drastically reduce the number of 
polar cosets that we need to evaluate in our algorithm. 
This complexity reduction makes it possible to 
compute the weight distribution of polar codes 
as a decreasing monomial codes at length 128. 

\subsection{Notations}
Here we specify some notation conventions 
we follow in this paper.  All the vectors in this paper are row vectors, 
unless otherwise specified. We use bold letters like 
$\bfu$ to denote vectors, and non-bold letters like $u_i$ 
to denote symbols within that vector. 
We let the indices for the symbols within vectors start from zero.
We use $\bfu_i$ to represent $(u_0,u_1,\cdots,u_i)$, 
a subvector of $\bfu$ with its first $(i+1)$ symbols.
We denote the concatenation of two vectors $\bfu$ and $\bfv$ as $(\bfu,\bfv)$.

\section{Polar Codes and Polar Cosets}
\label{sec:polar_cosets}
In this section, we briefly review polar codes, 
and give the definition for polar cosets, 
an essential concept in our work. 

Assuming $n = 2^m$, an $(n,k)$ polar code is a binary linear block 
code generated by $k$ rows in the polar transformation matrix 
$G_n = B_n K_2^{\otimes m}$, where $B_n$ is the 
bit-reversal permutation matrix, $K_2^{\otimes m}$ is the 
$m$-th Kronecker power of $K_2$, and 
\begin{displaymath}
K_2 = \begin{bmatrix}
1 & 0 \\ 1 & 1
\end{bmatrix}. 
\end{displaymath}

The encoding of polar codes is given by $\bfc = \bfu G_n$, 
where $\bfu$ is a length-$n$ binary input vector carrying $k$ data bits, 
and $\bfc$ is the codeword for transmission. 
The positions of the $k$ data bits in $\bfu$ 
are specified by an information index set $\cA$ 
of size $k$, with $\cA\subseteq\{0,1,\cdots,n-1\}$.
The remaining $n-k$ bits in $\bfu$ are set to 0, 
which are called frozen bits.
We also use $\cF = \{0,1,\cdots,n-1\}\backslash\cA$ to denote 
the frozen index set that specifies the 
positions of the frozen bits. 

We now give the definition for polar cosets. 
\begin{definition}
For a vector $\bfu_i\in\{0,1\}^{i+1}$ with $0\le i\le n-1$, 
we define the {\it polar coset} for path $\bfu_i$ as the affine space
\begin{displaymath}
C_n(\bfu_i) \;\deff\;
\left\{
(\bfu_i,\bfu')G_n
\;|\;
\bfu'\in\{0,1\}^{n-i-1}
\right\}
\end{displaymath}
where $(\bfu_i,\bfu')$ represents the concatenation of $\bfu$ and $\bfu'$, 
and $G_n$ is the polar transformation matrix. 
\label{def:polar_coset}
\end{definition}
\begin{example}
Consider polar transformation matrix 
$G_8$ with its rows denoted by $g_0,g_1,\cdots,g_7$ as shown 
in \Fref{fig:G8}.
\begin{figure}[!t]
\centering
\begin{displaymath}
\footnotesize
\arraycolsep=2.0pt\def\arraystretch{0.9}
G_8=
\left[
\begin{array}{cccc cccc}
1 & 0 & 0 & 0 & 0 & 0 & 0 & 0 \\
\rowcolor{LightCyan}
1 & 0 & 0 & 0 & 1 & 0 & 0 & 0 \\
1 & 0 & 1 & 0 & 0 & 0 & 0 & 0 \\
\rowcolor{LightCyan}
1 & 0 & 1 & 0 & 1 & 0 & 1 & 0 \\
1 & 1 & 0 & 0 & 0 & 0 & 0 & 0 \\
\rowcolor{Gray}
1 & 1 & 0 & 0 & 1 & 1 & 0 & 0 \\
\rowcolor{Gray}
1 & 1 & 1 & 1 & 0 & 0 & 0 & 0 \\
\rowcolor{Gray}
1 & 1 & 1 & 1 & 1 & 1 & 1 & 1 \\
\end{array}
\right]
\begin{array}{cccc cccc cccc cccc}
g_0\\ g_1\\ g_2\\ g_3\\ g_4\\ g_5\\ g_6\\ g_7
\end{array}
\end{displaymath}
\caption{Polar transformation matrix $G_8$ in Example \ref{ex:polar_coset}}
\label{fig:G8}
\end{figure}

Let $\bfu_4 = (0,1,0,1,0)$, then the polar coset $C_8(\bfu_4)$ is the 
affine space generated by $g_5,g_6,g_7$, and shifted by $g_1$ and $g_3$: 
\begin{displaymath}
C_8(\bfu_4) = g_1 + g_3 + \vspan\{g_5,g_6,g_7\}
\end{displaymath}
In \Fref{fig:G8}, those rows are highlighted in gray and in cyan, respectively.
\label{ex:polar_coset}
\end{example}

In this paper, we will mainly discuss the weight distribution 
of polar cosets, which can be described by their weight enumerating functions.
\begin{definition}
For a vector $\bfu_i\in\{0,1\}^{i+1}$ with $0\le i\le n-1$, 
we define the weight enumerating function 
for polar coset $C_n(\bfu_i)$ 
as the polynomial 
\begin{equation*}
A_n(\bfu_i)(X) \;\deff\; \sum_{w=0}^{n} A_w X^w ,
\end{equation*}
where $A_w$ is the number of vectors in $C_n(\bfu_i)$ with Hamming weight $w$. 
\label{def:coset_WEF}
\end{definition}

In prior works, the weight distribution of polar coset $C_n(\bfu_i)$, where 
$\bfu_i = (0,0,\cdots,0,1)$ is a length-$(i+1)$ all-zero 
vector with a single 1 at the end, is also referred to as {\it polar spectrum} 
in \cite{niu2019polar}, and as the {\it weight distribution for SC decoding of polar codes} 
in \cite{polyanskaya2020weight}. 
It has also been pointed out in \cite[Sec.III.B]{niu2019polar} 
and \cite[Sec.II.C]{polyanskaya2020weight} 
that the weight distribution of such polar coset $C_n(\bfu_i)$ can be used to 
analyze the error probability of the bit channels.

\section{Computing the Weight Enumerating Function of Polar Cosets}
\label{sec:WEF_algorithm}
In this section, we present the first key result of this paper: 
a recursive procedure that computes the weight enumerating function 
for arbitrary polar cosets. Recently, the authors in 
\cite{niu2019polar} and \cite{polyanskaya2020weight}
have introduced their respective algorithms 
that compute the weight distribution for polar coset 
$C_n(\bfu_i)$ with $\bfu_i=(0,0,\cdots,0,0)$ and 
$\bfu_i=(0,0,\cdots,0,1)$. 
However, how to efficiently compute the weight distribution for 
$C_n(\bfu_i)$ with arbitrary path $\bfu_i$ remains an open problem. 
Here we present a recursive computation procedure 
with time complexity $O(n^2)$ that solves this problem.

Let us first establish some notations. 
We use $\bfu_{\even}$ and $\bfu_{\odd}$ to 
denote the subvectors $(u_0,u_2,\cdots)$ and $(u_1,u_3,\cdots)$ 
of $\bfu$ with only even indices and only odd indices, respectively. 
We use $\bfu_{i,\even}$ and $\bfu_{i,\odd}$ to denote the subvectors 
of $\bfu_i$ with only even indices and only odd indices, respectively.

Our algorithm for polar cosets is based on the following recursive relations. 
\begin{theorem}
Let $m \ge 0$, $n = 2^m$, and $0\le i\le n-1$, then
\begin{multline}
A_{2n}(\bfu_{2i})(X) = \\
\sum_{u_{2i+1}\in\{0,1\}}
A_n(\bfu_{2i,\even} \oplus 
(\bfu_{2i,\odd}, u_{2i+1}))(X) \\
\cdot A_n(\bfu_{2i,\odd}, u_{2i+1})(X),
\label{eq:WEF_even}
\end{multline}
and 
\begin{multline}
A_{2n}(\bfu_{2i+1})(X) = \\
A_n(\bfu_{2i+1,\even} \oplus
\bfu_{2i+1,\odd}
)(X)
\cdot A_n(\bfu_{2i+1,\odd})(X).
\label{eq:WEF_odd}
\end{multline}
\label{thm:WEF_recursive}
\end{theorem}
\begin{IEEEproof}
Let $m \ge 0$ and $n=2^m$. 
For any $\bfu\in\{0,1\}^{2n}$, we have 
\begin{equation}
\begin{split}
\bfu \cdot G_{2n} 
&= (\bfu\cdot B_{2n}) K_2^{\otimes (m+1)} \\
&= (\bfu_{\even} \cdot B_n,\, \bfu_{\odd} \cdot B_n)
\begin{bmatrix}
K_2^{\otimes m} & 0 \\
K_2^{\otimes m} & K_2^{\otimes m} 
\end{bmatrix} \\
&= \Big((\bfu_{\even} \oplus \bfu_{\odd})\cdot 
B_nK_2^{\otimes m},\,
\bfu_{\odd}\cdot B_nK_2^{\otimes m}\Big) \\
&= \Big((\bfu_{\even} \oplus \bfu_{\odd})\cdot G_n,\,
\bfu_{\odd} \cdot G_n\Big)
\end{split}
\label{eq:uG2n}
\end{equation}
We first prove equation (\ref{eq:WEF_even}). 
According to Definition \ref{def:polar_coset}, we have
\begin{equation}
C_{2n}(\bfu_{2i}) = 
\left\{
(\bfu_{2i},\bfu')G_{2n} \;|\;
\bfu'\in\{0,1\}^{2n-2i-1}
\right\}
\label{eq:C2n}
\end{equation}
Let us represent $\bfu'$ as $\bfu' = (u_{2i+1},\bfv)$.
By looking at the two values $u_{2i+1}$ can take, 
$C_{2n}(\bfu_{2i})$ in (\ref{eq:C2n}) can be partitioned as:
\begin{equation}
\begin{split}
C_{2n}&(\bfu_{2i}) \\
&= \bigcup_{u_{2i+1}\in\{0,1\}}
C_{2n}(\bfu_{2i},u_{2i+1}) \\
&= \bigcup_{u_{2i+1}\in\{0,1\}}
\left\{
(\bfu_{2i},u_{2i+1},\bfv)G_{2n} \;|\;
\bfv\in\{0,1\}^{2n-2i-2}
\right\}
\end{split}
\label{eq:C2n_partition}
\end{equation}
Via~(\ref{eq:uG2n}), we can write 
$(\bfu_{2i},u_{2i+1},\bfv) G_{2n}$ 
in (\ref{eq:C2n_partition}) as 
\begin{multline}
(\bfu_{2i},u_{2i+1},\bfv) G_{2n} = \\
\Big(
(\bfu_{2i,\even} \oplus (\bfu_{2i,\odd}, u_{2i+1}),\, 
\bfv_{\even}\oplus \bfv_{\odd})\cdot G_{2n}, \\
(\bfu_{2i,\odd},u_{2i+1},
\bfv_{\odd})\cdot G_{2n}
\Big)
\end{multline}
Notice
when $\bfv$ ranges over $\{0,1\}^{2n-2i-2}$, 
both $\bfv_{\odd}$ and 
$(\bfv_{\even}\oplus\bfv_{\odd})$ 
range over $\{0,1\}^{n-i-1}$ separately. 
Thus we have
\begin{multline}
C_{2n}(\bfu_{2i},u_{2i+1}) = \\
\Big\{
(\bfc_1,\bfc_2) 
\;|\;
\bfc_1\in C_n(\bfu_{2i,\even} \oplus 
(\bfu_{2i,\odd}, u_{2i+1})),\, \\
\bfc_2\in C_n(\bfu_{2i,\odd}, u_{2i+1})
\Big\}
\label{eq:direct_sum}
\end{multline}
Hence for each $u_{2i+1}\in\{0,1\}$,
$C_{2n}(\bfu_{2i},u_{i+1})$ in (\ref{eq:C2n_partition}) 
can be expressed as the 
{\it direct sum} of 
two polar cosets 
\cite[\S 9 of Ch.\,2]{macwilliams1977theory}.
In other words, $C_{2n}(\bfu_{2i},u_{i+1})$ 
consists of all vectors $(\bfc_1,\bfc_2)$, 
where $\bfc_1\in C_n(\bfu_{2i,\even} \oplus (\bfu_{2i,\odd}, u_{2i+1})),$
and $\bfc_2\in C_n(\bfu_{2i,\odd}, u_{2i+1}).$

Since the weight enumerating function of 
the direct sum of two polar cosets equals the 
product of their two individual weight distribution functions, 
we obtain equation (\ref{eq:WEF_even}).
Equation (\ref{eq:WEF_odd}) follows in the 
same way by rewritting (\ref{eq:direct_sum}) as
\begin{multline}
C_{2n}(\bfu_{2i+1}) = \\
\Big\{
(\bfc_1,\bfc_2) 
\;|\;
\bfc_1\in C_n(\bfu_{2i+1,\even} \oplus 
\bfu_{2i+1,\odd}),\, \\
\bfc_2\in C_n(\bfu_{2i+1,\odd})
\Big\}  
\end{multline}
\end{IEEEproof}
\begin{figure*}[t!]
\centering
\begin{tikzpicture}[>=stealth,thick]

\def\la{1}
\def\h{0.8}
\def\gap{0.8}

\node [left] at (-0.3,0.0) {
		\begin{tabular}{r}
			$A_{n}(\bfu_{i-1},0)(X)\leftarrow$ \\[-0.1mm]
			$A_{n}(\bfu_{i-1},1)(X)\leftarrow$ \\
		\end{tabular}
        };
\node at (0,0) {\fbox{\mathstrut $A_{n}$}};

\draw [-] (\gap,0) -- (\gap+\la,0.6*\h);
\draw [-] (\gap,0) -- (\gap+\la,-0.6* \h);

\node at (2*\gap + \la, +\h) {\fbox{\mathstrut $A_{n/2}$}};
\node at (2*\gap + \la, -\h) {\fbox{\mathstrut $A_{n/2}$}};

\draw [-] (3*\gap + \la, \h) -- (3*\gap + 2*\la, 1.6*\h);
\draw [-] (3*\gap + \la, \h) -- (3*\gap + 2*\la, 0.4*\h);

\draw [-] (3*\gap + \la, -\h) -- (3*\gap + 2*\la, -1.6*\h);
\draw [-] (3*\gap + \la, -\h) -- (3*\gap + 2*\la, -0.4*\h);

\node at (4*\gap + 2*\la, 0) {$\cdots$};

\draw [-] (5*\gap + 2*\la, 2*\h) -- (5*\gap + 3*\la, 2.6*\h);
\draw [-] (5*\gap + 2*\la, 2*\h) -- (5*\gap + 3*\la, 1.4*\h);

\node [right] at (6*\gap + 3*\la+0.15, 2.6*\h) {
		\begin{tabular}{l}
			$\,\leftarrow 1$ \\[-0.5mm]
			$\,\leftarrow X$ \\
		\end{tabular}
		};
\node at (6*\gap + 3*\la, 2.6*\h) {\fbox{\mathstrut $A_{1}$}};

\node [right] at (6*\gap + 3*\la+0.15, 1.4*\h) {
		\begin{tabular}{l}
			$\,\leftarrow 1$ \\[-0.5mm]
			$\,\leftarrow X$ \\
		\end{tabular}
		};
\node at (6*\gap + 3*\la, 1.4*\h) {\fbox{\mathstrut $A_{1}$}};

\draw [-] (5*\gap + 2*\la, -2*\h) -- (5*\gap + 3*\la, -2.6*\h);
\draw [-] (5*\gap + 2*\la, -2*\h) -- (5*\gap + 3*\la, -1.4*\h);

\node [right] at (6*\gap + 3*\la+0.15, -2.6*\h) {
		\begin{tabular}{l}
			$\,\leftarrow 1$ \\[-0.5mm]
			$\,\leftarrow X$ \\
		\end{tabular}
		};
\node at (6*\gap + 3*\la, -2.6*\h) {\fbox{\mathstrut $A_{1}$}};
\node [right] at (6*\gap + 3*\la+0.15, -1.4*\h) {
		\begin{tabular}{l}
			$\,\leftarrow 1$ \\[-0.5mm]
			$\,\leftarrow X$ \\
		\end{tabular}
		};
\node at (6*\gap + 3*\la, -1.4*\h) {\fbox{\mathstrut $A_{1}$}};

\node at (6*\gap + 3*\la, 0) {$\vdots$};

\end{tikzpicture}
\caption{The recursive procedure that computes the weight enumerating 
function for polar cosets}
\label{fig:recursive_procedure}
\end{figure*}

In \Tref{thm:WEF_recursive}, equation (\ref{eq:WEF_even}) and 
equation (\ref{eq:WEF_odd}) can also be written as
\begin{multline}
A_{2n}(\bfu_{2i-1},u_{2i})(X) = \\
\sum_{u_{2i+1}\in\{0,1\}}
A_n(\bfu_{2i-1,\even} \oplus \bfu_{2i-1,\odd}
\,,\,
u_{2i}\oplus u_{2i+1})(X) \\
\cdot A_n(\bfu_{2i-1,\odd},u_{2i+1})(X)
\label{eq:WEF_even_v2}
\end{multline}
and as
\begin{multline}
A_{2n}(\bfu_{2i},u_{2i+1})(X) = \\
A_n(\bfu_{2i-1,\even} \oplus
\bfu_{2i-1,\odd},\,
u_{2i} \oplus u_{2i+1}
)(X) \\
\cdot A_n(\bfu_{2i-1,\odd},u_{2i+1})(X),
\label{eq:WEF_odd_v2}
\end{multline}
respectively. 
In this way, equation (\ref{eq:WEF_even_v2}) 
and equation (\ref{eq:WEF_odd_v2}) fall 
into forms similar to the recursive relations for the bit channels 
\cite[Equations (22) and (23)]{arikan2009channel}. 
Therefore, similar to the recursive procedure that computes 
the probabilities for the bit channels,  
we can also compute the weight enumerating functions of 
polar cosets recursively with the stopping conditions:
\begin{equation}
    A_1(0) = 1, \quad A_1(1) = X.
\end{equation}
This recursive procedure is illustrated in Figure 
\ref{fig:recursive_procedure},
and its steps are shown in Algorithm \ref{alg:CalcA}. 
\begin{algorithm}[t!]
\caption{CalcA($n,\bfu_{i-1}$)}
\label{alg:CalcA}
	\KwIn{block length $n$ and vector $\bfu_{i-1}$}
	\KwOut{
	    a pair of polynomials \\
	    \phantom{\textbf{Output:}}
    	$\left(A_n(\bfu_{i-1},0)(X),
	    A_n(\bfu_{i-1},1)(X)\right)$
	}
	\eIf(\tcp*[f]{Stopping conditions}){$n = 1$}
	{
	    \Return{$(1,X)$}
	}
	{
	    \eIf
	    {$i\mod 2 = 0$}
	    {
	        $(f_0,f_1) \leftarrow$ CalcA($n/2,\bfu_{i-1,\even} \oplus
            \bfu_{i-1,\odd}$) \\
	        $(g_0,g_1) \leftarrow$ CalcA($n/2,\bfu_{i-1,\odd}$)  \\
	        \Return{$(f_0g_0 + f_1g_1, f_0g_1 + f_1g_0)$} \\
	        \hfill\tcp{Use (\ref{eq:WEF_even})} 
	    }
	    {
	        $(f_0,f_1) \leftarrow$ CalcA($n/2,\bfu_{i-2,\even} \oplus
            \bfu_{i-2,\odd}$) \\
	        $(g_0,g_1) \leftarrow$ CalcA($n/2,\bfu_{i-2,\odd}$)  \\
	        \eIf{$u_{i-1} = 0$}
	        {
	            \Return{$(f_0g_0,f_1g_1)$}
	            \hfill\tcp{Use (\ref{eq:WEF_odd})} 
	        }
	        {
	            \Return{$(f_1g_0,f_0g_1)$}
	            \hfill\tcp{Use (\ref{eq:WEF_odd})} 
	        }
	    }
	}
\end{algorithm}

We make the following remarks 
for Algorithm \ref{alg:CalcA}:
\begin{itemize}
    \item
    The object for recursion in Algorithm \ref{alg:CalcA} is a 
    pair of weight enumerating functions 
    $A_n(\bfu_{i-1},0)(X)$ and $A_n(\bfu_{i-1},1)(X)$. 
    \item 
    If we want to compute the weight distribution for polar coset 
    $C_n(\bfu_i)$, we should run Algorithm \ref{alg:CalcA} 
    with inputs $n$ and $\bfu_{i-1}$, 
    and select one of the two weight enumerating functions 
    from the output corresponding to the desired $u_i$. 
\end{itemize}
Next, we prove that Algorithm \ref{alg:CalcA} has time complexity $O(n^2)$.
\begin{theorem}
Algorithm \ref{alg:CalcA} has time complexity $O(n^2)$.
\end{theorem}
\begin{IEEEproof}
In Algorithm \ref{alg:CalcA}, depending on the inputs $i$ and $\bfu_{i-1}$, 
we have the following three cases for the lines we need to run:
\begin{enumerate}
\item[] \hspace{-9mm}Case 1: 
	\hspace{0mm}When $i$ is even, we run lines 5, 6 and 7.
\item[] \hspace{-9mm}Case 2: 
	\hspace{0mm}When $i$ is odd and $u_i=0$, we run lines 9, 10 and 12.
\item[] \hspace{-9mm}Case 3: 
	\hspace{0mm}When $i$ is odd and $u_i=1$, we run lines 9, 10 and 14.
\end{enumerate}

First, the complexity of line 5 is the same as that of line 9, 
and the complexity of line 6 is the same as that of line 10. 
Then for line 7, we need to do 4 polynomial multiplications and 2 polynomial additions, 
while for line 12 or line 14, we only need to do 2 polynomial multiplications. 
So case 1 has the highest complexity among the above three cases. 
Thus, henceforth we only consider case 1. 

Denote by $T(n)$ the time complexity of Algorithm \ref{alg:CalcA}.  
For the recursive part in the algorithm, 
line 5 and line 6 both take time $T(n/2)$.
For the non-recursive part, in line 7 we need to do 
4 polynomial multiplications and 2 polynomial additions. 
Since $f_0,f_1,g_0,g_1$ are weight enumerating functions 
of polar cosets with block length $n/2$, 
all of them have degrees at most $n/2$. 
Assume multiplication of two degree-$n$ 
polynomials takes time $O(n^2)$, and addition of two degree-$n$ 
polynomials takes time $O(n)$. It follows that
\begin{displaymath}
T(n) \le 2T(n/2) + 4\cdot O(n^2/4) + 2\cdot O(n), 
\end{displaymath}
which by the Master theorem \cite{cormen2009introduction} 
gives us $T(n)=O(n^2)$. 
\end{IEEEproof}

We also remark that the time complexity of 
Algorithm \ref{alg:CalcA} 
may be improved assuming multiplication of 
two degree-$n$ polynomials takes time $O(n\log n)$ 
with the Fast-Fourier Transform.

\section{Computing the Entire Weight Distribution of Polar Codes}
\label{sec:polar_WD}
In this section, we present a deterministic algorithm that computes 
the entire weight distribution of polar codes. 
We first show that any polar code can be represented as a disjoint 
union of certain polar cosets. 
This allows us to obtain the weight distribution of the 
entire code by summing up the weight distributions of those polar cosets. 
However, the number of polar cosets in this representation 
scales exponentially with a new parameter that we introduce herein, 
called the mixing factor.
We also show that our approach naturally 
extends to polar codes with dynamically frozen bits. 

\subsection{Representing Polar Codes with Polar Cosets}
First, we introduce two new parameters of polar codes 
that we call the last frozen index and the mixing factor, respectively.
\begin{definition}
Consider an $(n,k)$ polar code $\mbC$ 
specified in terms of its information index set~$\cA$.
With $\cF = \{0,1,\dots,n-1\}\backslash\cA$, 
we define the {\it last frozen index} of $\mbC$ as 
\begin{displaymath}
\tau(\mbC) 
\,\deff\,
\max\{\cF\}, 
\end{displaymath}
and define the {\it mixing factor} of $\mbC$ as
\begin{displaymath}
\MF(\mbC) 
\,\deff\,
\bigl| \left\{ i \in \cA ~|~ i < \tau(\mbC) \right\} \bigr|.
\end{displaymath}
\label{def:lfi_mf}
\end{definition}

Loosely speaking, the mixing factor of $\mbC$ 
counts the number of information bits appear before the last frozen bit.
It is easy to see that $\MF(\mbC)$ can be computed from $\tau(\mbC)$ 
as follows:
\begin{equation}
\begin{split}
\MF(\mbC) 
&\,=\,
k - \bigl| \left\{ i \in \cA ~|~ i > \tau(\mbC) \right\} \bigr| \\
&\,=\,
\tau(\mbC) - (n\,{-}\,k) + 1
\end{split}
\label{eq:MF-tau}
\end{equation}

Starting with an example, we now show that any polar code 
can be represented as a disjoint union of polar cosets.
\begin{example}
In this example, we denote the (16,11,4) extended Hamming code as $\mbC_H$.
It can be generated by rows in the polar transformation matrix $G_{16}$. 
Thus we can view $\mbC_H$ as a polar code of length 16 
with frozen index set $\cF=\{0,1,2,4,8\}$.
The polar transformation matrix $G_{16}$ is shown in \Fref{fig:G16}. 
\begin{figure}[t]
\begin{equation*}
{
\footnotesize
\arraycolsep=2.0pt\def\arraystretch{0.9}
\begin{array}{c}
u_0\\
u_1\\
u_2\\
{\color{red}u_3}\\
u_4\\
{\color{red}u_5}\\
{\color{red}u_6}\\
{\color{red}u_7}\\
u_8\\
{\color{blue}u_9}\\
{\color{blue}u_{10}}\\
{\color{blue}u_{11}}\\
{\color{blue}u_{12}}\\
{\color{blue}u_{13}}\\
{\color{blue}u_{14}}\\
{\color{blue}u_{15}}\\
\end{array}
\left[
\begin{array}{cccc cccc cccc cccc}
1 & 0 & 0 & 0 & 0 & 0 & 0 & 0 & 0 & 0 & 0 & 0 & 0 & 0 & 0 & 0 \\
1 & 0 & 0 & 0 & 0 & 0 & 0 & 0 & 1 & 0 & 0 & 0 & 0 & 0 & 0 & 0 \\
1 & 0 & 0 & 0 & 1 & 0 & 0 & 0 & 0 & 0 & 0 & 0 & 0 & 0 & 0 & 0 \\
1 & 0 & 0 & 0 & 1 & 0 & 0 & 0 & 1 & 0 & 0 & 0 & 1 & 0 & 0 & 0 \\
1 & 0 & 1 & 0 & 0 & 0 & 0 & 0 & 0 & 0 & 0 & 0 & 0 & 0 & 0 & 0 \\
1 & 0 & 1 & 0 & 0 & 0 & 0 & 0 & 1 & 0 & 1 & 0 & 0 & 0 & 0 & 0 \\
1 & 0 & 1 & 0 & 1 & 0 & 1 & 0 & 0 & 0 & 0 & 0 & 0 & 0 & 0 & 0 \\
1 & 0 & 1 & 0 & 1 & 0 & 1 & 0 & 1 & 0 & 1 & 0 & 1 & 0 & 1 & 0 \\
1 & 1 & 0 & 0 & 0 & 0 & 0 & 0 & 0 & 0 & 0 & 0 & 0 & 0 & 0 & 0 \\
1 & 1 & 0 & 0 & 0 & 0 & 0 & 0 & 1 & 1 & 0 & 0 & 0 & 0 & 0 & 0 \\
1 & 1 & 0 & 0 & 1 & 1 & 0 & 0 & 0 & 0 & 0 & 0 & 0 & 0 & 0 & 0 \\
1 & 1 & 0 & 0 & 1 & 1 & 0 & 0 & 1 & 1 & 0 & 0 & 1 & 1 & 0 & 0 \\
1 & 1 & 1 & 1 & 0 & 0 & 0 & 0 & 0 & 0 & 0 & 0 & 0 & 0 & 0 & 0 \\
1 & 1 & 1 & 1 & 0 & 0 & 0 & 0 & 1 & 1 & 1 & 1 & 0 & 0 & 0 & 0 \\
1 & 1 & 1 & 1 & 1 & 1 & 1 & 1 & 0 & 0 & 0 & 0 & 0 & 0 & 0 & 0 \\
1 & 1 & 1 & 1 & 1 & 1 & 1 & 1 & 1 & 1 & 1 & 1 & 1 & 1 & 1 & 1 \\
\end{array}
\right]
}
\end{equation*}
\caption{Polar transformation matrix $G_{16}$ 
in Example \ref{ex:Hamming}}
\label{fig:G16}
\end{figure}

In \Fref{fig:G16}, the information bits of $\mbC_H$ are 
highlighted in red and blue, and the frozen bits are black.
We color the information bits appearing before the last frozen bit 
in red, and color the rest of the information bits in blue.
The last frozen bit of $\mbC_H$ is $u_8$, 
so the last frozen index of $\mbC_H$ is $\tau(\mbC_H)=8$. 
The mixing factor of $\mbC_H$ counts the number of red bits, 
so the mixing factor of $\mbC_H$ is $\MF(\mbC_H)=4$. 

Consider the polar coset $C_{16}(\bfu_8)$. 
For any binary vector $\bfu_8$ with $u_0=u_1=u_2=u_4=u_8=0$, 
and $u_3,u_5,u_6,u_7\in\{0,1\}$, the polar coset $C_{16}(\bfu_8)$ 
will be a subset of $\mbC_H$. In total we have $2^4 = 16$ options 
to assign the values for $u_3,u_5,u_6,u_7$. 
Hence there are $16$ such disjoint polar cosets, 
and the union of them is the entire code $\mbC_H$:
\begin{displaymath}
\mbC_H = \bigcup_{
\bfu_8\in\{0,1\}^9:\;
u_0=u_1=u_2=u_4=u_8=0
}
C_{16}(\bfu_8)
\end{displaymath}
Therefore, the entire weight distribution of $\mbC_H$ 
can be obtained by first computing the weight enumerating functions 
of all those 16 polar cosets, and then taking the sum.
\label{ex:Hamming}
\end{example}

This polar coset representation for general polar codes 
can be summarized by the following proposition.
\begin{proposition}
Let $\mbC$ be a polar code with frozen index set $\cF$, 
and last frozen index $\tau$.
Then $\mbC$ can be represented as a disjoint union of polar cosets as: 
\begin{displaymath}
\mbC = \bigcup_{\substack{
\bfu_{\tau}\in\{0,1\}^{\tau+1}:\;
u_i=0\text{ for all }i\in\cF
}}
C_{n}(\bfu_{\tau})
\end{displaymath}
The number of polar cosets in this representation equals $2^{\MF(\mbC)}$.
\label{pp:polar_WEF}
\end{proposition}

\subsection{Representing Polar Codes with Dynamically Frozen Bits}
We now show that our polar coset representation in 
\Pref{pp:polar_WEF} extends to polar codes with {\it dynamically frozen bits}. 
Polar codes with dynamically frozen bits, 
first introduced in \cite{trifonov2013dynamic}, 
are polar codes where each of the frozen bits $u_i$ 
is not necessarily fixed to be zero, 
but can be set as a linear function of its previous bits as $u_i = f_i(\bfu_{i-1})$. 
For frozen bits with indices in $\cF$, 
we refer to those boolean functions $\{f_i~|~i\in\cF\}$ 
as the {\it dynamic constraints} for the code.
Examples of polar codes with dynamically frozen bits are 
polar codes with CRC precoding \cite{tal2015list}, 
polar subcodes \cite{trifonov2015polar},
polarization-adjusted convolutional (PAC) codes \cite{arikan2019PAC}, etc.
In fact, since any binary linear code can be represented 
as a polar code with dynamically frozen bits
\cite{trifonov2013dynamic}, our representation extends to 
all binary linear codes, as well. 

The concept of last frozen index and mixing factor 
in Definition \ref{def:lfi_mf} naturally extends 
to polar codes with dynamically frozen bits.
We again illustrate our polar coset representation 
with an example, in which the Hamming code in Example \ref{ex:Hamming} 
is slightly modified so its frozen bits become dynamically frozen.

\begin{example}
Denote by $\mbC_H'$ a $(16,11)$ polar code with 
frozen index set $\cF=\{0,1,2,4,8\}$, 
where $u_0,u_1,u_2$ are frozen as 0, 
and $u_4$ and $u_8$ are dynamically frozen as 
$u_4=u_3$ and $u_8=u_5+u_6$, respectively.
We have $\tau(\mbC_H')=8$ and $\MF(\mbC_H')=4$ 
the same as in Example \ref{ex:Hamming}.
\begin{figure}[t]
\begin{equation*}
{
\footnotesize
\arraycolsep=2.0pt\def\arraystretch{0.9}
\begin{array}{l}
u_0\\
u_1\\
u_2\\
{\color{red}u_3}\\
u_4 = u_3\\
{\color{red}u_5}\\
{\color{red}u_6}\\
{\color{red}u_7}\\
u_8 = u_5 + u_6\\
{\color{blue}u_9}\\
{\color{blue}u_{10}}\\
{\color{blue}u_{11}}\\
{\color{blue}u_{12}}\\
{\color{blue}u_{13}}\\
{\color{blue}u_{14}}\\
{\color{blue}u_{15}}\\
\end{array}
\left[
\begin{array}{cccc cccc cccc cccc}
1 & 0 & 0 & 0 & 0 & 0 & 0 & 0 & 0 & 0 & 0 & 0 & 0 & 0 & 0 & 0 \\
1 & 0 & 0 & 0 & 0 & 0 & 0 & 0 & 1 & 0 & 0 & 0 & 0 & 0 & 0 & 0 \\
1 & 0 & 0 & 0 & 1 & 0 & 0 & 0 & 0 & 0 & 0 & 0 & 0 & 0 & 0 & 0 \\
1 & 0 & 0 & 0 & 1 & 0 & 0 & 0 & 1 & 0 & 0 & 0 & 1 & 0 & 0 & 0 \\
1 & 0 & 1 & 0 & 0 & 0 & 0 & 0 & 0 & 0 & 0 & 0 & 0 & 0 & 0 & 0 \\
1 & 0 & 1 & 0 & 0 & 0 & 0 & 0 & 1 & 0 & 1 & 0 & 0 & 0 & 0 & 0 \\
1 & 0 & 1 & 0 & 1 & 0 & 1 & 0 & 0 & 0 & 0 & 0 & 0 & 0 & 0 & 0 \\
1 & 0 & 1 & 0 & 1 & 0 & 1 & 0 & 1 & 0 & 1 & 0 & 1 & 0 & 1 & 0 \\
1 & 1 & 0 & 0 & 0 & 0 & 0 & 0 & 0 & 0 & 0 & 0 & 0 & 0 & 0 & 0 \\
1 & 1 & 0 & 0 & 0 & 0 & 0 & 0 & 1 & 1 & 0 & 0 & 0 & 0 & 0 & 0 \\
1 & 1 & 0 & 0 & 1 & 1 & 0 & 0 & 0 & 0 & 0 & 0 & 0 & 0 & 0 & 0 \\
1 & 1 & 0 & 0 & 1 & 1 & 0 & 0 & 1 & 1 & 0 & 0 & 1 & 1 & 0 & 0 \\
1 & 1 & 1 & 1 & 0 & 0 & 0 & 0 & 0 & 0 & 0 & 0 & 0 & 0 & 0 & 0 \\
1 & 1 & 1 & 1 & 0 & 0 & 0 & 0 & 1 & 1 & 1 & 1 & 0 & 0 & 0 & 0 \\
1 & 1 & 1 & 1 & 1 & 1 & 1 & 1 & 0 & 0 & 0 & 0 & 0 & 0 & 0 & 0 \\
1 & 1 & 1 & 1 & 1 & 1 & 1 & 1 & 1 & 1 & 1 & 1 & 1 & 1 & 1 & 1 \\
\end{array}
\right]
}
\end{equation*}
\caption{Polar transformation matrix $G_{16}$ 
in Example \ref{ex:Hamming_dyn}}
\label{fig:G16_dyn}
\end{figure}

Consider the polar coset $C_{16}(\bfu_8)$. 
For any binary vector $\bfu_8$ with $u_3,u_5,u_6,u_7\in\{0,1\}$, 
if we let $u_0=u_1=u_2=0$, $u_4=u_3$ and $u_8=u_5+u_6$, 
then the polar coset $C_{16}(\bfu_8)$ will be a subset of $\mbC_H'$. 
Thus similar to Example \ref{ex:Hamming}, with $2^4 = 16$ options 
to assign the values for $u_3,u_5,u_6$ and $u_7$, 
$\mbC_H'$ can be represented as a disjoint union of 16 disjoint polar cosets as
\begin{displaymath}
\mathbb{C} = \bigcup_{\substack{
\bfu_8\in\{0,1\}^9:\;
u_0=u_1=u_2=0,\,u_4=u_3, \\
u_8=u_5+u_6
}}
C_{16}(\bfu_8)
\end{displaymath}
\label{ex:Hamming_dyn}
\end{example}

In general, Proposition \ref{pp:polar_WEF} 
extends to polar codes with dynamically frozen bits 
as follows.
\begin{proposition}
Let $\mbC$ be a polar code with dynamically frozen bits, 
with frozen index set $\cF$, last frozen index $\tau$, 
and the dynamic constraints $\{f_i~|~i\in\cF\}$. 
Then $\mbC$ can be represented as a disjoint 
union of polar cosets as: 
\begin{displaymath}
\mbC = 
\bigcup_{
\bfu_{\tau}\in\{0,1\}^{\tau+1}:\;
u_i=f_i(\bfu_{i-1})\text{ for all }i\in\cF
}
C_{n}(\bfu_\tau)
\end{displaymath}
The number of polar cosets in this representation equals $2^{\MF(\mbC)}$.
\label{pp:polar_dyn_WEF}
\end{proposition}

\subsection{Computing the Entire Weight Distribution}
This polar coset representation directly gives us a way to compute the 
weight distribution of polar codes. We can compute the weight enumerating 
function of each polar coset in the representation 
using Algorithm \ref{alg:CalcA}, and then take their sum to obtain the 
weight distribution of the entire code.
This procedure is shown in Algorithm \ref{alg:WEF_polar_dyn}, 
in which conventional polar codes are considered 
as special cases of polar codes with dynamically frozen bits. 
\begin{algorithm}[t!]
\caption{Compute the weight enumerating functions of polar codes 
with dynamically frozen bits}
\label{alg:WEF_polar_dyn}
	\KwIn{
	block length $n$, frozen index set $\cF$, and 
	dynamic constraint $\{f_i~|~i\in\cF\}$
	}
	\KwOut{weight enumerating function $A_{\mathbb{C}}(X)$ 
	of polar code $\mbC$}
	$\tau\leftarrow\max\{\cF\}$\\
	$A_{\mathbb{C}}(X)\leftarrow 0$ \\
	\For{
	$\bfu_{\tau}\in\{0,1\}^{\tau+1}:\;
	u_i=f_i(\bfu_{i-1})\text{ for all }i\in\cF$
	}
	{
	    \tcp{Use Algorithm \ref{alg:CalcA}}
	    $(f_0,f_1) \leftarrow$ CalcA($n,\bfu_{\tau-1}$) \\
		$u_\tau\leftarrow f_{\tau}(\bfu_{\tau-1})$ \\
	    \eIf{$u_\tau=0$}
	    {
	        $A_{\mathbb{C}}(X) \leftarrow A_{\mathbb{C}}(X) + f_0$ 
	    }
	    {
	        $A_{\mathbb{C}}(X) \leftarrow A_{\mathbb{C}}(X) + f_1$ 
	    }
    }
    \Return{$A_{\mbC}(X)$}
\end{algorithm}

For a polar code $\mbC$, the number of polar cosets 
in the representation equals 2 to the power of $\MF(\mbC)$. 
For each polar coset, both the computation of its 
dynamically frozen bits and the computation of its 
weight enumerating function via Algorithm 1 have complexity $O(n^2)$.
Thus without parallel computation, 
Algorithm \ref{alg:WEF_polar_dyn} has time complexity 
$O(2^{\MF(\mbC)}\,n^2)$. 
It is clear that this complexity is largely governed 
by the mixing factor of the code. 
For reference, we list the mixing factors of 
several rate 1/2 polar codes from length 8 to length 1024 in Table \ref{tb:mf_5G}.
Those polar codes are constructed using the reliability sequence
in the 5G technical specification \cite{3GPP}.
\begin{table}[!t]
\renewcommand{\arraystretch}{1.3}
\caption{Mixing factor of rate 1/2 polar codes constructed 
using the reliability sequence in 5G \cite{3GPP}}
\label{tb:mf_5G}
\centering
\begin{tabular}{|c|c|c|c|c|c|c|c|c|}
\hline
code length $n$ & 8 & 16 & 32 & 64 & 128  & 256  & 512  & 1024 \\
\hline
$\MF(\mbC)$ & 1 & 2 & 9 & 17 & 34 & 73 & 161 & 385 \\
\hline
\end{tabular}
\end{table}

Unfortunately, this approach turns out to be inefficient
for polar codes with CRCs. For a polar code concatenated 
with a CRC outer code \cite{tal2015list}, 
since all the CRC parity bits are located at the end 
of the data vector, the mixing factor of the code 
would be the same as the code dimension. 
In this case, Algorithm~\ref{alg:WEF_polar_dyn} will have complexity 
higher than that of the brute-force search. 

For PAC codes \cite{arikan2019PAC}, 
their mixing factors are determined by the rate profiles.
For PAC codes with polar rate profiles, 
their mixing factors will be the same as polar codes.
For PAC codes with Reed-Muller rate profiles, 
which show better performance under sequential decoding 
and list decoding \cite{arikan2019PAC,yao2021list,rowshan2021polarization}, 
their mixing factors will be the same as Reed-Muller codes. 
As will be shown in Section~\ref{sec:mixing_factor}, 
Reed-Muller codes have relatively larger mixing factors 
compared with polar codes.

We also list the mixing factors of several 
extended BCH codes represented as polar codes with 
dynamically frozen bits in Table~\ref{tb:mf_eBCH}.
Those codes are obtained by extending some of the primitive narrow-sense 
BCH codes listed in Table A-1 in \cite[Appendix A]{clark2013error}. 
Note that for a given binary linear code, 
its representations as polar codes with dynamically frozen 
bits will be different for different codeword bit orders.
Since it is known that primitive BCH codes contain as subcodes 
punctured Reed-Muller code of the same designed distance 
\cite[Ch. 13. \S5. Theorem 11]{macwilliams1977theory}, 
we permute the bit positions of those extended BCH codes 
from the {\it cyclic order} to the {\it standard order} 
\cite{kasami1993complexity}, such that heuristically, 
their representations as polar codes 
with dynamically frozen bits have smaller mixing factors.
This standard order is also used in \cite{trifonov2015polar} 
to construct polar subcodes from extended BCH codes.
In Table~\ref{tb:mf_eBCH}, we also specify the extended BCH codes 
that are equivalent to Reed-Muller codes RM($r$,$m$) of order $r$ and length $n=2^m$. 

Note that the weight distribution of extended BCH codes at length 128 
have already been computed by Desaki, Fujiwara and Kasami in \cite{desaki1997weight}.
Here, we only list those mixing factors as a reference.
Compared with polar codes, we can see that the mixing factors 
for extended BCH codes that have large code distances are also much larger, 
which indicates that our approach is less applicable here. 
\begin{table}[!t]
\renewcommand{\arraystretch}{1.3}
\caption{Mixing factors of extended BCH (EBCH) 
codes as polar codes with dynamically frozen bits.
The distances of the codes are obtained from \cite{morelos2006art} and from \cite{desaki1997weight}.} 
\label{tb:mf_eBCH}
\centering
\begin{tabular}{|l|c|c|}
\hline
EBCH($n$,\,$k$) & distance $d$ & mixing factor $\MF(\mbC)$ \\ \hline
EBCH(8,\,4) = RM(1,\,3) & 4 & 1 \\ 
EBCH(16,\,7) & 6 & 4 \\ 
EBCH(32,\,16) = RM(2,\,5) & 8 & 9 \\ 
EBCH(64,\,30) & 14 & 23 \\ 
EBCH(64,\,36) & 12 & 29 \\ 
EBCH(128,\,57) & 24 & 50 \\ 
EBCH(128,\,64) & 22 & 57 \\ 
\hline
\end{tabular}
\end{table}

\section{Mixing Factor of Polar Codes}
\label{sec:mixing_factor}
Note that the approach described in Section \ref{sec:polar_WD} 
applies to polar codes in a general setting where: 
(1) the frozen bits can be dynamically frozen;
(2) the information index set can be arbitrary. 
Hereforth, we focus on conventional polar codes where: 
(1) the frozen bits are all frozen to zero; 
(2) for code of dimension $k$, 
the information index set $\cA$ is chosen 
such that the corresponding bit channels 
are the ``best'' $k$ bit channels. 
In \Arikan{}'s definition, the $k$ bit channels 
with the smallest Bhattacharyya parameters are selected. 
Two alternative criteria for picking the best $k$ bit channels 
are mutual information and error probabilities.
If we follow either one of these three criteria, 
polar codes fall into the category of 
decreasing monomial codes, 
first introduced in \cite{bardet2016algebraic}. 

In this section, we briefly review 
the definition of decreasing monomial codes. 
Then, to upper bound the complexity of 
Algorithm \ref{alg:WEF_polar_dyn}, 
we prove that self-dual Reed-Muller codes 
have the largest mixing factors 
among all decreasing monomial codes 
with rates at most $1/2$.

\subsection{Decreasing Monomial Codes}
We start by reviewing the definition of monomial codes.
Let $n=2^m$, and let the polynomial ring given by
\begin{multline*}
\cR_m = \mathbb{F}[x_0,x_1,\cdots,x_{m-1}] \\
/(x_0^2-x_0,x_1^2-x_1,\cdots,x_{m-1}^2-x_{m-1}). 
\end{multline*}
Each polynomial $p\in\mathcal{R}_m$ can be associated with 
a binary vector in $\mathbb{F}_2^{n}$ as the evaluation of $p$ 
in all the binary entries 
$\bfx=(x_0,\cdots,x_{m-1})\in\mathbb{F}_2^m$. 
In other words, polynomial $p$ is associated with 
$\ev(p)=(p(\bfx))_{\bfx\in\mathbb{F}_2^m}$
where $\ev:\cR_m\rightarrow\mathbb{F}_2^n$ is a homomorphism from the 
polynomials to the binary $n$-tuples.
In this work, we specify the order of $\bfx$ in vector 
$(p(\bfx))_{\bfx\in\mathbb{F}_2^m}$
such that from left to right, 
the number $\sum_{i=0}^{m-1}z_i2^i$ is in ascending order 
from 0 to $2^m-1$, where the binary vector 
$(z_0,z_1,\cdots,z_{m-1})$ is defined by: 
\begin{displaymath}
(z_0,z_1,\cdots,z_{m-1}) = (1-x_{m-1},1-x_{m-2},\cdots,1-x_0)
\end{displaymath}
Denote the set of all the monomials in $\cR_m$ as
\begin{equation*}
\mathcal{M}_m  = \left\{
x_0^{b_0}x_1^{b_1} \cdots x_{m-1}^{b_{m-1}} \;\Big|\;
(b_0,b_1,\cdots,b_{m-1})\in\mathbb{F}_2^m
\right\}.
\end{equation*}
The monomial codes can be defined as follows. 
\begin{definition}
Let $n=2^m$ and $\cI\in\cM_m$, the monomial code $\mathbb{C}(\cI)$ 
generated by $\cI$ is the linear space 
\begin{equation*}
    \mbC(\cI) \;\deff\; \vspan\{\ev(f)\:|\:f\in \cI\}.
\end{equation*}
\end{definition}

Since every row in the polar transformation matrix $G_n$ 
can be obtained as $\ev(f)$ with some $f\in\cM_m$, 
polar codes can be viewed as monomial codes.
For a monomial $f \in \cM_m$ given by
$f = x_{i_1}x_{i_2} \cdots x_{i_d}$,
we write:
\begin{align*}
    \deg f      &= d, \\
    \ind(f)     &= \{i_1,i_2,\ldots,i_d\}, \\
    [f]         &= (a_{m-1},a_{m-2},\ldots,a_0) \in \{0,1\}^m, \\
    \rowind{f}  &= \sum_{i=0}^{m-1} \!a_i 2^i 
                = \sum_{i=0}^{m-1} (1{-}b_i)2^i,
\end{align*}
where the two binary vectors 
$(a_{m-1},a_{m-2},\ldots,a_0)$ and $(b_{m-1},b_{m-2},\ldots,b_0)$
are defined by:
\begin{displaymath}
f = x_0^{1-a_0} x_1^{1-a_1} \cdots x_{m-1}^{1-a_{m-1}}
  = x_0^{b_0} x_1^{b_1} \cdots x_{m-1}^{b_{m-1}}
\end{displaymath}
Following this notation, if we label the rows 
in the polar transformation matrix $G_n$ with indices from 0 to $n-1$, 
the evaluation $\ev(f)$ for a monomial $f\in\cM_m$ 
has row index $\rowind{f}$ in $G_n$, 
and $[f]$ contains the digits in the binary expansion of $\rowind{f}$.
When the underlying $G_n$ is clear from the context, 
we simply refer $\rowind{f}$ as the {\it row index} for $f$. 
\begin{example}
Consider the polar transformation matrix $G_{16}$. 
The monomials in $\cM_4$ whose evaluations are rows
in $G_{16}$ are shown in Figure \ref{fig:mono}.
\begin{figure}[t]
\begin{equation*}
{
\footnotesize
\arraycolsep=1.8pt\def\arraystretch{1.0}
\raisebox{10pt}{
$
\begin{array}{ccc}
f            & [f]       & \rowind{f}\\
&&\\
x_0x_1x_2x_3 & (0,0,0,0) & 0\\
x_1x_2x_3    & (0,0,0,1) & 1\\
x_0x_2x_3    & (0,0,1,0) & 2\\
x_2x_3       & (0,0,1,1) & 3\\
x_0x_1x_3    & (0,1,0,0) & 4\\
x_1x_3       & (0,1,0,1) & 5\\
x_0x_3       & (0,1,1,0) & 6\\
x_3          & (0,1,1,1) & 7\\
x_0x_1x_2    & (1,0,0,0) & 8\\
x_1x_2       & (1,0,0,1) & 9\\
x_0x_2       & (1,0,1,0) & 10\\
x_2          & (1,0,1,1) & 11\\
x_0x_1       & (1,1,0,0) & 12\\
x_1          & (1,1,0,1) & 13\\
x_0          & (1,1,1,0) & 14\\
1            & (1,1,1,1) & 15\\
\end{array}
$
}
\left[
\begin{array}{cccc cccc cccc cccc}
1 & 0 & 0 & 0 & 0 & 0 & 0 & 0 & 0 & 0 & 0 & 0 & 0 & 0 & 0 & 0 \\
1 & 0 & 0 & 0 & 0 & 0 & 0 & 0 & 1 & 0 & 0 & 0 & 0 & 0 & 0 & 0 \\
1 & 0 & 0 & 0 & 1 & 0 & 0 & 0 & 0 & 0 & 0 & 0 & 0 & 0 & 0 & 0 \\
1 & 0 & 0 & 0 & 1 & 0 & 0 & 0 & 1 & 0 & 0 & 0 & 1 & 0 & 0 & 0 \\
1 & 0 & 1 & 0 & 0 & 0 & 0 & 0 & 0 & 0 & 0 & 0 & 0 & 0 & 0 & 0 \\
1 & 0 & 1 & 0 & 0 & 0 & 0 & 0 & 1 & 0 & 1 & 0 & 0 & 0 & 0 & 0 \\
1 & 0 & 1 & 0 & 1 & 0 & 1 & 0 & 0 & 0 & 0 & 0 & 0 & 0 & 0 & 0 \\
1 & 0 & 1 & 0 & 1 & 0 & 1 & 0 & 1 & 0 & 1 & 0 & 1 & 0 & 1 & 0 \\
1 & 1 & 0 & 0 & 0 & 0 & 0 & 0 & 0 & 0 & 0 & 0 & 0 & 0 & 0 & 0 \\
1 & 1 & 0 & 0 & 0 & 0 & 0 & 0 & 1 & 1 & 0 & 0 & 0 & 0 & 0 & 0 \\
1 & 1 & 0 & 0 & 1 & 1 & 0 & 0 & 0 & 0 & 0 & 0 & 0 & 0 & 0 & 0 \\
1 & 1 & 0 & 0 & 1 & 1 & 0 & 0 & 1 & 1 & 0 & 0 & 1 & 1 & 0 & 0 \\
1 & 1 & 1 & 1 & 0 & 0 & 0 & 0 & 0 & 0 & 0 & 0 & 0 & 0 & 0 & 0 \\
1 & 1 & 1 & 1 & 0 & 0 & 0 & 0 & 1 & 1 & 1 & 1 & 0 & 0 & 0 & 0 \\
1 & 1 & 1 & 1 & 1 & 1 & 1 & 1 & 0 & 0 & 0 & 0 & 0 & 0 & 0 & 0 \\
1 & 1 & 1 & 1 & 1 & 1 & 1 & 1 & 1 & 1 & 1 & 1 & 1 & 1 & 1 & 1 \\
\end{array}
\right]
}
\end{equation*}
\caption{Polar transformation matrix $G_{16}$ 
in Example \ref{ex:mono}}
\label{fig:mono}
\end{figure}
\label{ex:mono}
\end{example}

Henceforth, whenever we write a monomial as $f = x_{i_1}x_{i_2} \cdots x_{i_d}$
we assume that $i_1 < i_2 < \cdots < i_d$, unless stated otherwise.
A partial order on the monomials in $\cM_m$ is 
introduced in \cite{bardet2016algebraic} as follows:
\begin{definition}
\label{BDOT-order}
If $f = x_{i_1}x_{i_2} \cdots x_{i_d}$ and $g = x_{j_1}x_{j_2} \cdots x_{j_d}$ 
are two monomials of the same degree $d$, 
we write $f \preceq g$ if
\begin{equation*}
i_1 \le j_1, \quad i_2 \le j_2, \quad\cdots,\quad i_d \le j_d
\end{equation*}
If $\deg f < \deg g$, we 
write $f \preceq g$ if there exists a divisor 
$g^*$ of $g$, 
such that $g^*$ has the same degree as $f$ and $f \preceq g^*$.
If $f \preceq g$ and $f\ne g$, we write $f \prec g$.
\end{definition}

It has been shown by Bardet, Dragoi, Otmani, 
and Tillich in \cite{bardet2016algebraic}, 
and by Sch\"urch in \cite{schurch2016partial} that 
polar codes satisfy the following property.
\begin{theorem}
\label{BDOT-degradation}
Let $\mbC_n(\cA)$ be a polar code,
specified in terms of its information index set $\cA$. 
If $\rowind{g}\in\cA$ and $f \preceq g$, 
then also $\rowind{f}\in\cA$. Equivalently, 
if $\rowind{f}\in\cF$ and $f \preceq g$, then also
$\rowind{g}\in\cF$.
\end{theorem}

Therefore, the authors in \cite{bardet2016algebraic} 
call the family of all codes having this property 
as decreasing monomial codes.
\begin{definition}[Decreasing monomial codes 
\cite{bardet2016algebraic}]
Decreasing monomial codes inclues all monomial codes that 
satisfy Theorem~\ref{BDOT-degradation}.
\end{definition}

Besides polar codes, the family of decreasing monomial codes 
also includes Reed-Muller codes.
A simple lemma about the partial order of two monomials, 
and their row indices that is easy to verify is the following:
\begin{lemma}
If $g \preceq f$, then $\rowind{g} \ge \rowind{f}$.
\label{lm:total_order_is_a_linear_extension}
\end{lemma}

\subsection{The Largest Mixing Factor of Polar Codes}
Now we are ready to study the range of 
mixing factor of decreasing monomial codes. 
Since by the MacWilliams identity \cite{MacWilliams}, 
one can easily obtain the weight distribution of 
a code from the weight distribution of its dual, 
if we want to compute the weight distribution of 
a given decreasing monomial code, we have the options of 
applying Algorithm 2 to either the code itself, or to its dual. 
On the other hand, Bardet, Dragoi, Otmani, 
and Tillich have shown that the dual of any decreasing monomial code 
is also a decreasing monomial code \cite[Proposition 6]{bardet2016algebraic}. 
Thus to get a complexity cap of our approach, 
it suffices to limit our space to 
decreasing monomial codes of rates at most 1/2.
\begin{theorem}
Let $\mbC$ be an $(n,k)$ decreasing monomial code 
with $n=2^m$, $m=2t+1$, and dimension $k \le n/2$, 
then 
\begin{equation}
\MF(\mbC) \le 2^{2t} - 2^{t+1} + 1
\label{eq:thm4}
\end{equation}
Moreover, the equality holds only when 
$\mbC$ is the self-dual Reed-Muller code.
\label{thm:RM2t+1}
\end{theorem}

According to Theorem \ref{thm:RM2t+1}, the mixing factor 
of decreasing monomial codes at length $n=2^m$, 
where $m$ is an odd number, 
is bounded by the mixing factor of self-dual Reed-Muller codes.
Here we list the mixing factor of self-dual Reed-Muller codes at length 
8, 32, 128, 512 and 2048 in Table \ref{tb:mf_RM}.
\begin{table}[!t]
\renewcommand{\arraystretch}{1.3}
\caption{Mixing factors of self-dual Reed-Muller codes}
\label{tb:mf_RM}
\centering
\begin{tabular}{|c|c|c|c|c|c|c|c|c|}
\hline
code length $n$
& 8 & 32 & 128  & 512 & 2048\\ \hline
mixing factor 
& 1 & 9 & 49 & 225 & 961 \\ \hline
\end{tabular}
\end{table}

For decreasing monomial codes at length $n=2^m$, 
where $m$ is an even number, we make the following conjecture 
about their largest mixing factors based on numerical observation.
The conjectured upper bounds for decreasing monomial codes at length 
16, 64, 256, 1024 are listed in Table \ref{tb:mf_conjecture}.
\begin{conjecture}
Let $\mbC$ be an $(n,k)$ decreasing monomial code, 
with $n = 2^m$, $m=2t$, and dimension $k \le n/2$, then 
$$\MF(\mbC) \le 2^{2t-1} - 2^{t+1} + 2$$
where the equality is achievable.
\label{thm:RM2t}
\end{conjecture}
\begin{table}[!t]
\renewcommand{\arraystretch}{1.3}
\caption{Conjectured upper bounds for decreasing monomial codes 
with rates $\le$ 1/2}
\label{tb:mf_conjecture}
\centering
\begin{tabular}{|c|c|c|c|c|c|c|c|c|}
\hline
code length $n$
& 16 & 64 & 256 & 1024 \\ \hline
mixing factor $\le$
& 2 & 18 & 98 & 450 \\ \hline
\end{tabular}
\end{table}

The rest of this section is devoted to the proof 
of Theorem \ref{thm:RM2t+1}.
\begin{table}[!t]
\renewcommand{\arraystretch}{1.3}
\caption{A table illustrating the positions of $g$ and $g'$ 
in the proof of \Tref{thm:RM2t+1}}
\label{tb:M2t+1}
\centering
\begin{tabular}{|c|c|c|}
    \hline
    $\rowind{f}$ & $[f]$ & $f$ \\
    \hline
    0 & $(\underbrace{0,0,\cdots,0,0}_{2t+1})$ & $x_0x_1x_2\cdots x_{2t}$ \\
    \hline
    1 & $(0,0,\cdots,0,1)$ & $x_1x_2\cdots x_{2t}$ \\ 
    \hline
    \vdots & \vdots & \vdots \\ 
    \hline
    $2^{2t+1}-2^{t+1}$ & 
    $(\underbrace{1,\cdots,1}_{t},\underbrace{0,\cdots,0,0}_{t+1})$ 
    & {\color{blue}$g=x_0x_1x_2\cdots x_{t}$} \\
    \hline
    $2^{2t+1}-2^{t+1}+1$ & 
    $(\underbrace{1,\cdots,1}_{t},0,\cdots,0,1)$ 
    & {\color{blue}$g'=x_1x_2\cdots x_{t}$} \\
    \hline
    \vdots & \vdots & \vdots \\ 
    \hline
    $2^{2t+1}-2$ & $(1,1,\cdots,1,0)$ & $x_0$ \\ \hline
    $2^{2t+1}-1$ & $(1,1,\cdots,1,1)$ & $1$ \\ \hline
\end{tabular}
\end{table}
\begin{IEEEproof}[Proof of Theorem \ref{thm:RM2t+1}]
It can be verified by exhaustive search 
that Theorem \ref{thm:RM2t+1} holds when $t=1$ and $t=2$. 
So hereforth, we focus on proving the theorem when $t\ge 3$.
In this proof we use Table \ref{tb:M2t+1} to help illustrate our arguments.
First we show self-dual Reed-Muller codes achieve the equality 
in (\ref{eq:thm4}).
\begin{quotation}
\noindent
\textbf{Claim\,1.}
Let $\mbC$ be the self-dual Reed-Muller code of length $2^{2t+1}$, 
then $\MF(\mbC) = 2^{2t} - 2^{t+1} + 1$.

\vspace{0.7ex}
\textit{Proof.} 
Let $\cI$ be set of monomials generating $\mbC$, 
Then $\cI$ contains all monomials of degree less or equal to $t$. 
Referring to Table \ref{tb:M2t+1}, we have 
$\tau(\mbC) = 2^{2t+1} - 2^{t+1}$. 
Thus from equation (\ref{eq:MF-tau}), we have
\begin{equation*}
    \MF(\mbC) = 2^{2t} - 2^{t+1} + 1. 
\end{equation*}
\end{quotation}
Then we focus on the following claim, which states that if the mixing 
factor of the code is at least $2^{2t} - 2^{t+1} + 1$, 
then the code has to be the self-dual Reed-Muller code. 
\begin{quotation}
\noindent
\textbf{Claim\,2.}
Let $\mbC$ be a decreasing monomial code of length $n = 2^{2t+1}$ 
and dimension $k \le n/2$. 
If $$\MF(\mbC) \ge 2^{2t} - 2^{t+1} + 1,$$
then $\mbC$ can only be the self-dual Reed-Muller code.
\end{quotation}
Now it suffices to prove Claim 2, since it is clear that 
\Tref{thm:RM2t+1} follows if we combine Claim 1 and Claim 2.
Hereforth, we denote $g = x_0x_1x_2\cdots x_t$ as the monomial with 
$\rowind{g}=2^{2t+1} - 2^{t+1}$, and denote $g' = x_1x_2\cdots x_t$ 
as the monomial with $\rowind{g'}=\rowind{g} + 1$. 
If we list out all the monomials in $\cM_{2t+1}$ following their row indices, 
the positions of $g$ and $g'$ in this list are shown in Table \ref{tb:M2t+1}. 

Our proof for Claim 2 relies on the following three claims.
\begin{quotation}
\noindent
\textbf{Claim\,3.}
Let $\mbC$ be a decreasing monomial code of length $n = 2^{2t+1}$, 
and frozen index set $\cF$. 
If $\tau(\mbC)\ge \rowind{g}$, then $\rowind{g}\in\cF$.

\vspace{0.7ex}
\textit{Proof.} 
Observe from Table \ref{tb:M2t+1} that for any monomial $h$ with 
$\rowind{h}\ge\rowind{g}$, $h$ is a divisor of $g$, 
which gives us $h\preceq g$. 
Therefore, if $\tau(\mbC)\ge \rowind{g}$, 
it follows from \Tref{BDOT-degradation} that $\rowind{g}\in\cF$. 

\vspace{2.0ex}
\noindent
\textbf{Claim\,4.}
Let $\mbC$ be a decreasing monomial code of length $n = 2^{2t+1}$, and frozen index set $\cF$. 
If $\tau(\mbC)\ge \rowind{g'}$, then $\rowind{g'}\in\cF$.

\vspace{0.7ex}
\textit{Proof.} 
It can be observed from Table \ref{tb:M2t+1} that, for any monomial $h$ with 
$\rowind{h}\ge\rowind{g'}$, we have $h\preceq g'$. 
Therefore, 
similar to the proof for Claim 3, 
if $\tau(\mbC)\ge \rowind{g'}$, 
it follows from \Tref{BDOT-degradation} that $\rowind{g'}\in\cF$. 

\vspace{2.0ex}
\noindent
\textbf{Claim\,5.}
Let $\mbC$ be a decreasing monomial code of length $n = 2^{2t+1}$ and dimension $k \le n/2$.
If 
$$
\MF(\mbC) \ge 2^{2t} - 2^{t+1} + 1
\quad\text{and}\quad
\tau(\mbC) = \rowind{g}
,$$
then $\mbC$ is the self-dual Reed-Muller code. 

\vspace{0.7ex}
\textit{Proof.} 
Since for any monomial $h$ with $\deg h\ge t+1$, we have $g\preceq h$, 
it follows from \Tref{BDOT-degradation} that $\rowind{h}\in\cF$ 
for any monomial $h$ with degree at least $t+1$. 
So $\mbC$ is a subcode of the self-dual Reed-Muller code.
On the other hand, in view of equation (\ref{eq:MF-tau}), 
the dimension of the code is at least
$$k = \MF(\mbC) + (n-1) - \tau(\mbC) \ge n/2$$
Thus $\mbC$ can only be the self-dual Reed-Muller code itself.
\end{quotation}
At this point, we are ready to prove Claim 2. 
We will first show that given the conditions in Claim 2, 
$g$ has to be frozen. Moreover, we will then show that 
the last frozen index of the code has to be exactly $\rowind{g}$.
\begin{quotation}
\noindent
\textit{Proof of Claim 2.} 
First from equation (\ref{eq:MF-tau}), we have
\begin{equation*}
    \tau(\mbC) = \MF(\mbC) + (n-k) - 1 \ge 
    2^{2t+1} - 2^{t+1}
\end{equation*}
So the last frozen index of $\mbC$ is at least $\rowind{g}$. 
Then we show that $\tau(\mbC) > \rowind{g}$ leads to a contradiction.
Assuming $\tau(\mbC) > \rowind{g}$, 
we have $\rowind{g}\in\cF$ and $\rowind{g'}\in\cF$ 
following Claim 3 and Claim 4, respectively.
Now we count the number of monomials having 
row indices in $\cF$.
First for any $h$ with $\deg h\ge t+1$, we have $h\succeq g$. 
Thus it follows from \Tref{BDOT-degradation} that $\rowind{h}\in\cF$ 
for all $h$ with $\deg h\ge t+1$. 
The number of those monomials can be counted as
$$
\sum_{i=t+1}^{2t+1}\binom{2t+1}{i} = 2^{2t}
$$
Then for any degree-$t$ monomial $h$ without $x_0$, 
we have $h\succeq g'$, which gives us $\rowind{h}\in\cF$ 
following \Tref{BDOT-degradation}. 
The number of those monomials can be counted as $\binom{2t}{t}$. 
Therefore, the number of frozen indices of $\mbC$ is at least
$$
|\cF| \ge 2^{2t} +  
\binom{2t}{t}
$$
This gives
$$
|\cA| = n - |\cF| \le 2^{2t} - 
\binom{2t}{t}
$$
But that contradicts $\MF(\mbC) \ge 2^{2t} - 2^{t+1} + 1$, since
$$
2^{2t} - 
\binom{2t}{t}
<
2^{2t} - 2^{t+1} + 1
$$
for all $t\ge 3$. 

Since $\tau(\mbC) > \rowind{g}$ leads to a contradiction, 
we can only have $\tau(\mbC) = \rowind{g}$. 
Thus it follows from Claim 5 that $\mbC$ can only be the 
self-dual Reed-Muller code.
\end{quotation}
\end{IEEEproof}

\section{Reducing Computation Complexity using a Subgroup of LTA}
\label{sec:LTA}
As a family of codes including polar codes, 
decreasing monomial codes have a large automorphism group. 
It was first shown that the automorphism group of 
decreasing monomial codes includes the lower triangular affine group (LTA) 
in \cite{bardet2016algebraic}.
Recently, this result has been extended to 
the block lower triangular affine group (BLTA) 
\cite{geiselhart2021automorphism}. 
In this section, we look into 
the algebraic properties of decreasing monomial codes, 
and focus on a subgroup of LTA.
We prove that this subgroup 
acts transitively on certain 
subsets of decreasing monomial codes.
This result implies that those subsets 
share the same weight distribution, 
allowing us to drastically reduce the complexity of our approach.

\subsection{Lower Triangular Affine Groups and Their Group Action}
We start by reviewing the definition for the lower triangular affine group, 
and how it acts on polynomials. Henceforth, 
binary $m\times m$ matrices are denoted by $\mbF_2^{m\times m}$, 
and $m$-tuples in $\mbF_2^m$ are treated as column vectors. 
Following the notation in \cite{bardet2016algebraic}, 
we denote the affine transformation $\bfx\mapsto A\bfx+\bfb$ 
over $\mbF_2^m$ by a pair $(A,\bfb)$, 
where $A\in\mbF_2^{m\times m}$ and $\bfb\in\mbF_2^m$.
\begin{definition}
The {\it lower triangular affine group}, 
denoted as $\LTA(m,2)$, 
consists of all affine transformations $(A,\bfb)$, 
where $A\in\mbF_2^{m\times m}$ 
is a non-singular lower triangular square matrix, 
and $\bfb\in\mathbb{F}_2^{m}$. 
\label{def:LTA}
\end{definition}

The group action of $\LTA(m,2)$ on the polynomial ring
$\cR_m$ can be defined as follows.
For an affine transformation $(A,\bfb)\in\LTA(m,2)$ with $A=(a_{ij})$, 
and a polynomial $p\in\cR_m$, 
we denote by $(A,\bfb)\cdot p$ the action of $(A,\bfb)$ on $p$, 
where each monomial $x_i$ in $p$ is replaced 
by another monomial $y_i$ defined as 
$$
y_i = \sum_{j=0}^{m-1}a_{ij}x_j + b_i.
$$

For monomials, the following expansion after the group action of 
$\LTA(m,2)$ can be observed, which follows directly from 
Definition~\ref{def:LTA}.
\begin{proposition}
Let $(A,\bfb)\in\LTA(m,2)$ and $f\in\cM_m$, 
then $(A,\bfb)\cdot f$ can be expanded as 
\begin{equation}
(A,\bfb)\cdot f = f + \sum_{g\in\cM_m:\; g\prec f} u_g\cdot g,
\label{eq:Abf}
\end{equation}
where $u_g\in\{0,1\}$ for all $g$. 
\label{pp:Abf}
\end{proposition}

Here is another way to view the action by the affine transformations.
Recall that the evaluation $\ev(p)$ of a polynomial $p\in\cR_m$ is 
a vector that consists of $p(\bfx)$ over all $\bfx\in\mbF_2^m$.
Since every affine transformation $(A,\bfb)$ is a bijection on $\mbF_2^m$, 
the evaluation $\ev((A,\bfb)\cdot p)$ 
can be obtained from $\ev(p)$ by permuting its coordinates. 
Denote the action of $(A,\bfb)$ on a polynomial evaluation as
$$(A,\bfb)\cdot \ev(p) = \ev((A,\bfb)\cdot p),$$
we can then view this action as a permutation 
on the coordinates of $\ev(p)$. 
In particular, vector $(A,\bfb)\cdot \ev(p)$ and 
vector $\ev(p)$ have the same Hamming weight.

In the work by Bardet, Dragoi, Otmani, and Tillich,  
they show that the automorphism group of 
decreasing monomial codes over $m$ variables 
includes the lower triangular affine group $\LTA(m,2)$
\cite[Theorem 2]{bardet2016algebraic}. 

\subsection{A Subgroup of $\LTA(m,2)$}
In the main theorem of this section, we consider a subgroup of $\LTA(m,2)$, 
denoted $\LTA(m,2)_f$, that we associate with a given monomial $f$. 
This subgroup was introduced in \cite{bardet2016algebraic}, 
where it was used to analyze and count the minimum weight codewords 
of decreasing monomial codes. 
\begin{definition}
Let $f\in\cM_m$. The subgroup $\LTA(m,2)_f$ of $\LTA(m,2)$ associated with the 
monomial $f$ is defined as
$$
\LTA(m,2)_f \;\deff\;
\{ (A,\bfb)\in\LTA(m,2)\;|\; A\in M_f, \bfb\in B_f \},
$$
where
\begin{multline*}
M_f = \{ (a_{ij})\in\mbF_2^{m\times m} \;|\;
\forall\;i>j,\\
a_{ij} = 0 \;\;\text{if}\;\;
i\not\in\ind(f)  \;\;\text{or}\;\; j\in\ind(f) \}
\end{multline*}
and
\begin{displaymath}
B_f = \{ \bfb\in \mbF_2^m \;|\; b_i=0
\;\;\text{if}\;\; i\not\in\ind(f) \}
\end{displaymath}
\label{def:LTA_f}
\end{definition}
\begin{example}
Consider $\LTA(5,2)$. 
For $f=x_0x_3x_4\in\cM_5$, we have
\begin{multline*}
M_f =  \{ (a_{ij})\in\mbF_2^{5\times 5} \;|\;
\forall\;i>j,\\ 
a_{ij} = 0 \;\;\text{if}\;\;
i\not\in\{0,3,4\} \;\;\text{or}\;\; j\in\{0,3,4\} \}
\end{multline*}
and
\begin{displaymath}
B_f = \{ \bfb\in \mbF_2^m \;|\; b_i=0
\;\;\text{if}\;\; i\not\in\{0,3,4\} \}
\end{displaymath}
Therefore, the affine transformations $(A,\bfb)$ 
in the subgroup $\LTA(5,2)_f$ have the form:
\begin{equation*}
    A = \begin{pmatrix}
    1 & 0 & 0 & 0 & 0\\
    0 & 1 & 0 & 0 & 0\\
    0 & 0 & 1 & 0 & 0\\
    0 & a_{3,1} & a_{3,2} & 1 & 0 \\
    0 & a_{4,1} & a_{4,2} & 0 & 1
    \end{pmatrix},
    \quad\text{ and }\quad
    b = \begin{pmatrix}
    b_0\\
    0\\
    0\\
    b_3\\
    b_4
    \end{pmatrix}, 
\end{equation*}
where $a_{3,1},\, a_{3,2},\, a_{4,1},\, a_{4,2},\, b_0,\, b_3,\, b_4$ 
can take any value in $\{0,1\}$. 
There are $2^7$ such affine transformations, 
so the order of $\LTA(2,4)_f$ is $2^7=128$.
\end{example}

\subsection{One-Variable Descendance Relation}
We also introduce a new relation on the monomials 
for our main theorem of this section. 
Henceforth, whenever we write $f = gx_i$ for two monomials $f$ and $g$, 
we assume $i\not\in\ind(g)$.
\begin{definition}
For $f,g\in\mathcal{M}_m$, we say 
$g$ is a {\it one-variable descendant} of $f$, 
and write $g\ovd f$ if either one of the following holds:
\begin{enumerate}
    \item $f = hx_i$ and $g = hx_j$ for some monomial $h$ with $j < i$.
    \item $f = gx_i$
\end{enumerate}
\label{def:ovd}
\end{definition}

Compared with the partial order in Definition \ref{BDOT-order}, 
this one-variable descendance relation is a more restricted relation 
in the sense that, the two involved monomials can only differ by {\it one} variable.
We remark that this one-variable descendance relation 
is only a relation, but not a partial order on the monomials. 
The following example shows that this new relation is not transitive.
\begin{example}
For monomials in $\cM_4$, we have
$$
x_0x_2 \ovd x_0x_1x_2,
\quad\text{and}\quad
x_0x_1x_2 \ovd x_0x_1x_3,
$$
but $x_0x_2$ is not a one-variable descendant of $x_0x_1x_3$.
\end{example}

\subsection{The Main Theorem: A Transitive Group Action}
Now we are ready to present the main theorem of this section. 
\begin{figure*}[t!]
\begin{equation*}
\footnotesize
{
\arraycolsep=2.0pt\def\arraystretch{0.9}
\begin{array}{r}
\begin{array}{c}
\phantom{u_0}\\ \phantom{u_1}\\ \phantom{u_2}\\ \phantom{u_3}\\ \phantom{u_4}\\ \phantom{u_5}\\
f\rightarrow\,\, \\
\end{array} \\
\cS\left\{\!\!
\begin{array}{c}
\rightarrow\\
\phantom{u_8}\\
\phantom{u_9}\\
\rightarrow\\
\phantom{u_{11}}\\
\rightarrow\\
\phantom{u_{13}}\\
\rightarrow\\
\end{array}
\right.\!\!\\
\begin{array}{c}
\phantom{u_{15}}\\
\phantom{u_{16}}\\
\phantom{u_{17}}\\
\phantom{u_{18}}\\
\phantom{u_{19}}\\
\phantom{u_{20}}\\
\phantom{u_{21}}\\
\phantom{u_{22}}\\
\phantom{u_{23}}\\
\phantom{u_{24}}\\
\phantom{u_{25}}\\
\phantom{u_{26}}\\
\phantom{u_{27}}\\
\phantom{u_{28}}\\
\phantom{u_{29}}\\
\phantom{u_{30}}\\
\phantom{u_{31}}
\end{array}
\end{array}
\hspace{-2mm}
\begin{array}{cccc cccc cccc cccc}
x_0x_1x_2x_3x_4\\
x_1x_2x_3x_4\\
x_0x_2x_3x_4\\
x_2x_3x_4\\
x_0x_1x_3x_4\\
x_1x_3x_4\\
{\framebox{\color{red}$x_0x_3x_4$}}\\
{\color{red}x_3x_4}\\
x_0x_1x_2x_4\\
{\color{orange}x_1x_2x_4}\\
{\color{red}x_0x_2x_4}\\
{\color{orange}x_2x_4}\\
{\color{red}x_0x_1x_4}\\
{\color{orange}x_1x_4}\\
{\color{red}x_0x_4}\\
{\color{orange}x_4}\\
x_0x_1x_2x_3\\
{\color{blue}x_1x_2x_3}\\
{\color{blue}x_0x_2x_3}\\
{\color{blue}x_2x_3}\\
{\color{blue}x_0x_1x_3}\\
{\color{blue}x_1x_3}\\
{\color{blue}x_0x_3}\\
{\color{blue}x_3}\\
{\color{blue}x_0x_1x_2}\\
{\color{blue}x_1x_2}\\
{\color{blue}x_0x_2}\\
{\color{blue}x_2}\\
{\color{blue}x_0x_1}\\
{\color{blue}x_1}\\
{\color{blue}x_0}\\
{\color{blue}1}
\end{array}
\begin{array}{c}
u_0\\
u_1\\
u_2\\
u_3\\
u_4\\
u_5\\
{\color{red}u_6}\\
{\color{red}u_7}\\
u_8\\
{\color{red}u_9}\\
{\color{red}u_{10}}\\
{\color{red}u_{11}}\\
{\color{red}u_{12}}\\
{\color{red}u_{13}}\\
{\color{red}u_{14}}\\
{\color{red}u_{15}}\\
u_{16}\\
{\color{blue}u_{17}}\\
{\color{blue}u_{18}}\\
{\color{blue}u_{19}}\\
{\color{blue}u_{20}}\\
{\color{blue}u_{21}}\\
{\color{blue}u_{22}}\\
{\color{blue}u_{23}}\\
{\color{blue}u_{24}}\\
{\color{blue}u_{25}}\\
{\color{blue}u_{26}}\\
{\color{blue}u_{27}}\\
{\color{blue}u_{28}}\\
{\color{blue}u_{29}}\\
{\color{blue}u_{30}}\\
{\color{blue}u_{31}}\\
\end{array}
\left[
\begin{array}{cccc cccc cccc cccc cccc cccc cccc cccc}
1 & 0 & 0 & 0 & 0 & 0 & 0 & 0 & 0 & 0 & 0 & 0 & 0 & 0 & 0 & 0 & 0 & 0 & 0 & 0 & 0 & 0 & 0 & 0 & 0 & 0 & 0 & 0 & 0 & 0 & 0 & 0\\
1 & 0 & 0 & 0 & 0 & 0 & 0 & 0 & 0 & 0 & 0 & 0 & 0 & 0 & 0 & 0 & 1 & 0 & 0 & 0 & 0 & 0 & 0 & 0 & 0 & 0 & 0 & 0 & 0 & 0 & 0 & 0\\
1 & 0 & 0 & 0 & 0 & 0 & 0 & 0 & 1 & 0 & 0 & 0 & 0 & 0 & 0 & 0 & 0 & 0 & 0 & 0 & 0 & 0 & 0 & 0 & 0 & 0 & 0 & 0 & 0 & 0 & 0 & 0\\
1 & 0 & 0 & 0 & 0 & 0 & 0 & 0 & 1 & 0 & 0 & 0 & 0 & 0 & 0 & 0 & 1 & 0 & 0 & 0 & 0 & 0 & 0 & 0 & 1 & 0 & 0 & 0 & 0 & 0 & 0 & 0\\
1 & 0 & 0 & 0 & 1 & 0 & 0 & 0 & 0 & 0 & 0 & 0 & 0 & 0 & 0 & 0 & 0 & 0 & 0 & 0 & 0 & 0 & 0 & 0 & 0 & 0 & 0 & 0 & 0 & 0 & 0 & 0\\
1 & 0 & 0 & 0 & 1 & 0 & 0 & 0 & 0 & 0 & 0 & 0 & 0 & 0 & 0 & 0 & 1 & 0 & 0 & 0 & 1 & 0 & 0 & 0 & 0 & 0 & 0 & 0 & 0 & 0 & 0 & 0\\
1 & 0 & 0 & 0 & 1 & 0 & 0 & 0 & 1 & 0 & 0 & 0 & 1 & 0 & 0 & 0 & 0 & 0 & 0 & 0 & 0 & 0 & 0 & 0 & 0 & 0 & 0 & 0 & 0 & 0 & 0 & 0\\
1 & 0 & 0 & 0 & 1 & 0 & 0 & 0 & 1 & 0 & 0 & 0 & 1 & 0 & 0 & 0 & 1 & 0 & 0 & 0 & 1 & 0 & 0 & 0 & 1 & 0 & 0 & 0 & 1 & 0 & 0 & 0\\
1 & 0 & 1 & 0 & 0 & 0 & 0 & 0 & 0 & 0 & 0 & 0 & 0 & 0 & 0 & 0 & 0 & 0 & 0 & 0 & 0 & 0 & 0 & 0 & 0 & 0 & 0 & 0 & 0 & 0 & 0 & 0\\
\rowcolor{Gray}
1 & 0 & 1 & 0 & 0 & 0 & 0 & 0 & 0 & 0 & 0 & 0 & 0 & 0 & 0 & 0 & 1 & 0 & 1 & 0 & 0 & 0 & 0 & 0 & 0 & 0 & 0 & 0 & 0 & 0 & 0 & 0\\
1 & 0 & 1 & 0 & 0 & 0 & 0 & 0 & 1 & 0 & 1 & 0 & 0 & 0 & 0 & 0 & 0 & 0 & 0 & 0 & 0 & 0 & 0 & 0 & 0 & 0 & 0 & 0 & 0 & 0 & 0 & 0\\
\rowcolor{Gray}
1 & 0 & 1 & 0 & 0 & 0 & 0 & 0 & 1 & 0 & 1 & 0 & 0 & 0 & 0 & 0 & 1 & 0 & 1 & 0 & 0 & 0 & 0 & 0 & 1 & 0 & 1 & 0 & 0 & 0 & 0 & 0\\
1 & 0 & 1 & 0 & 1 & 0 & 1 & 0 & 0 & 0 & 0 & 0 & 0 & 0 & 0 & 0 & 0 & 0 & 0 & 0 & 0 & 0 & 0 & 0 & 0 & 0 & 0 & 0 & 0 & 0 & 0 & 0\\
\rowcolor{Gray}
1 & 0 & 1 & 0 & 1 & 0 & 1 & 0 & 0 & 0 & 0 & 0 & 0 & 0 & 0 & 0 & 1 & 0 & 1 & 0 & 1 & 0 & 1 & 0 & 0 & 0 & 0 & 0 & 0 & 0 & 0 & 0\\
1 & 0 & 1 & 0 & 1 & 0 & 1 & 0 & 1 & 0 & 1 & 0 & 1 & 0 & 1 & 0 & 0 & 0 & 0 & 0 & 0 & 0 & 0 & 0 & 0 & 0 & 0 & 0 & 0 & 0 & 0 & 0\\
\rowcolor{Gray}
1 & 0 & 1 & 0 & 1 & 0 & 1 & 0 & 1 & 0 & 1 & 0 & 1 & 0 & 1 & 0 & 1 & 0 & 1 & 0 & 1 & 0 & 1 & 0 & 1 & 0 & 1 & 0 & 1 & 0 & 1 & 0\\
1 & 1 & 0 & 0 & 0 & 0 & 0 & 0 & 0 & 0 & 0 & 0 & 0 & 0 & 0 & 0 & 0 & 0 & 0 & 0 & 0 & 0 & 0 & 0 & 0 & 0 & 0 & 0 & 0 & 0 & 0 & 0\\
\rowcolor{Gray}
1 & 1 & 0 & 0 & 0 & 0 & 0 & 0 & 0 & 0 & 0 & 0 & 0 & 0 & 0 & 0 & 1 & 1 & 0 & 0 & 0 & 0 & 0 & 0 & 0 & 0 & 0 & 0 & 0 & 0 & 0 & 0\\
\rowcolor{Gray}
1 & 1 & 0 & 0 & 0 & 0 & 0 & 0 & 1 & 1 & 0 & 0 & 0 & 0 & 0 & 0 & 0 & 0 & 0 & 0 & 0 & 0 & 0 & 0 & 0 & 0 & 0 & 0 & 0 & 0 & 0 & 0\\
\rowcolor{Gray}
1 & 1 & 0 & 0 & 0 & 0 & 0 & 0 & 1 & 1 & 0 & 0 & 0 & 0 & 0 & 0 & 1 & 1 & 0 & 0 & 0 & 0 & 0 & 0 & 1 & 1 & 0 & 0 & 0 & 0 & 0 & 0\\
\rowcolor{Gray}
1 & 1 & 0 & 0 & 1 & 1 & 0 & 0 & 0 & 0 & 0 & 0 & 0 & 0 & 0 & 0 & 0 & 0 & 0 & 0 & 0 & 0 & 0 & 0 & 0 & 0 & 0 & 0 & 0 & 0 & 0 & 0\\
\rowcolor{Gray}
1 & 1 & 0 & 0 & 1 & 1 & 0 & 0 & 0 & 0 & 0 & 0 & 0 & 0 & 0 & 0 & 1 & 1 & 0 & 0 & 1 & 1 & 0 & 0 & 0 & 0 & 0 & 0 & 0 & 0 & 0 & 0\\
\rowcolor{Gray}
1 & 1 & 0 & 0 & 1 & 1 & 0 & 0 & 1 & 1 & 0 & 0 & 1 & 1 & 0 & 0 & 0 & 0 & 0 & 0 & 0 & 0 & 0 & 0 & 0 & 0 & 0 & 0 & 0 & 0 & 0 & 0\\
\rowcolor{Gray}
1 & 1 & 0 & 0 & 1 & 1 & 0 & 0 & 1 & 1 & 0 & 0 & 1 & 1 & 0 & 0 & 1 & 1 & 0 & 0 & 1 & 1 & 0 & 0 & 1 & 1 & 0 & 0 & 1 & 1 & 0 & 0\\
\rowcolor{Gray}
1 & 1 & 1 & 1 & 0 & 0 & 0 & 0 & 0 & 0 & 0 & 0 & 0 & 0 & 0 & 0 & 0 & 0 & 0 & 0 & 0 & 0 & 0 & 0 & 0 & 0 & 0 & 0 & 0 & 0 & 0 & 0\\
\rowcolor{Gray}
1 & 1 & 1 & 1 & 0 & 0 & 0 & 0 & 0 & 0 & 0 & 0 & 0 & 0 & 0 & 0 & 1 & 1 & 1 & 1 & 0 & 0 & 0 & 0 & 0 & 0 & 0 & 0 & 0 & 0 & 0 & 0\\
\rowcolor{Gray}
1 & 1 & 1 & 1 & 0 & 0 & 0 & 0 & 1 & 1 & 1 & 1 & 0 & 0 & 0 & 0 & 0 & 0 & 0 & 0 & 0 & 0 & 0 & 0 & 0 & 0 & 0 & 0 & 0 & 0 & 0 & 0\\
\rowcolor{Gray}
1 & 1 & 1 & 1 & 0 & 0 & 0 & 0 & 1 & 1 & 1 & 1 & 0 & 0 & 0 & 0 & 1 & 1 & 1 & 1 & 0 & 0 & 0 & 0 & 1 & 1 & 1 & 1 & 0 & 0 & 0 & 0\\
\rowcolor{Gray}
1 & 1 & 1 & 1 & 1 & 1 & 1 & 1 & 0 & 0 & 0 & 0 & 0 & 0 & 0 & 0 & 0 & 0 & 0 & 0 & 0 & 0 & 0 & 0 & 0 & 0 & 0 & 0 & 0 & 0 & 0 & 0\\
\rowcolor{Gray}
1 & 1 & 1 & 1 & 1 & 1 & 1 & 1 & 0 & 0 & 0 & 0 & 0 & 0 & 0 & 0 & 1 & 1 & 1 & 1 & 1 & 1 & 1 & 1 & 0 & 0 & 0 & 0 & 0 & 0 & 0 & 0\\
\rowcolor{Gray}
1 & 1 & 1 & 1 & 1 & 1 & 1 & 1 & 1 & 1 & 1 & 1 & 1 & 1 & 1 & 1 & 0 & 0 & 0 & 0 & 0 & 0 & 0 & 0 & 0 & 0 & 0 & 0 & 0 & 0 & 0 & 0\\
\rowcolor{Gray}
1 & 1 & 1 & 1 & 1 & 1 & 1 & 1 & 1 & 1 & 1 & 1 & 1 & 1 & 1 & 1 & 1 & 1 & 1 & 1 & 1 & 1 & 1 & 1 & 1 & 1 & 1 & 1 & 1 & 1 & 1 & 1\\
\end{array}
\right]
\hspace{-2mm}
\begin{array}{l}
\begin{array}{c}
\\ \\ \\ \\ \\ \\ \\ \\ \\
\end{array} \\
\left.
\begin{array}{c}
\\ \\ \\ \\ \\ \\ \\ \\ \\ \\ \\ \\ \\ \\ \\ \\ \\ \\ \\ \\ \\ \\ \\
\end{array}
\right\}\text{generate }\mbC(\cT)
\end{array}
}
\end{equation*}
\caption{Polar Transformation matrix $G_{32}$ in Example \ref{ex:G32}}
\label{fig:G32}
\end{figure*}

\begin{theorem}
Let $\mbC(\cI)$ be a decreasing monomial code generated by $\cI\in\cM_m$, 
and let $f$ be the monomial in $\cI$ with the smallest row index:
$$f = \underset{g\in\cI}{\mathrm{argmin}}\;\rowind{g}$$
We partition $\cI$ into the following disjoint union
$$\cI = \{f\} \,\cup\, \cS \,\cup\, \cT,$$
where $\cS$ consists of all one-variable descendant 
of $f$ with row indices smaller than $\tau(\mbC)$, 
and $\cT$ contains the rest of the monomials in $\cI$:
$$\cS = \{h\in\cI \;|\; h\ovd f \text{ and }[\![h]\!]<\tau(\mbC)\},$$
and
$$\cT = \cI\,\backslash \left(\, \{f\}\cup\cS \,\right).$$
Then the group action of subgroup $\LTA(m,2)_f$ on the set $\cX$ is transitive, 
where $\cX$ is the set consisting of cosets of $\mbC(\cT)$ defined as follows
\begin{multline}
\cX = \\
\left\{
\ev(f) + \sum_{h\in\cS} u_h\cdot \ev(h) + \mbC(\cT)
\,\Big|\, 
\forall\;h\in\cS,\;
u_h\in\{0,1\}
\right\}
\end{multline}
Therefore, all the cosets of $\mbC(\cT)$ in $\cX$ have the same weight distribution.
\label{thm:subset}
\end{theorem}

Before proving this theorem, we illustrate \Tref{thm:subset}
with an example, and show how we can use 
this theorem to reduce the complexity when computing 
the weight distribution of decreasing monomial codes.
\begin{example}
Consider a (32,24) decreasing monomial code $\mbC$ specified by the 
frozen index set $\cF=\{0,1,2,3,4,5,8,16\}$. 
The monomials corresponding to the rows 
in $G_{32}$ are shown in \Fref{fig:G32}, where
the information bits are highlighted in red, orange and blue, 
and the frozen bits are black. 
Code $\mbC$ has last frozen index $\tau(\mbC) = 16$, 
and mixing factor $\MF(\mbC) = 9$. 

\vspace{1mm}
\noindent \textbf{\textit{Illustrating \Tref{thm:subset}}}
\vspace{1mm}

Let $f = x_0x_3x_4$ be the monomial 
with the smallest row index in $\cI$. 
Then $\cI$ can be partitioned as
$$\cI = \{f\} \cup \cS \cup \cT,$$
where $\cS$ consists of four of 
the one-variable descendants of $f$
with row indices smaller than $\tau(\mbC) = 16$, 
and $\cT$ consists of the rest of the monomials in $\cI$:
\begin{align*}
\cS &= \{
{\color{red}x_3x_4},
{\color{red}x_0x_2x_4},
{\color{red}x_0x_1x_4},
{\color{red}x_0x_4}
\},\\
\cT &= \{
{\color{orange}x_1x_2x_4},
{\color{orange}x_2x_4},
{\color{orange}x_1x_4},
{\color{orange}x_4},
{\color{blue}x_1x_2x_3},
{\color{blue}x_0x_2x_3},
\cdots,
{\color{blue}x_0},
{\color{blue}1}
\}
\end{align*}
As shown in \Fref{fig:G32}, the monomials in $\cS$ are colored in red, 
the monomials in $\cT$ are colored in orange and in blue, 
and the subcode $\mbC(\cT)$ is generated by the gray rows in $G_{32}$. 

Then, set $\cX$ is defined to consist of 16 cosets of $\mbC(\cT)$ in the form
\begin{multline*}
\ev(f) + 
u_1\cdot\ev({\color{red}x_3x_4}) + 
u_2\cdot\ev({\color{red}x_0x_2x_4})\\ + 
u_3\cdot\ev({\color{red}x_0x_1x_4}) + 
u_4\cdot\ev({\color{red}x_0x_4}) + 
\mbC(\cT),
\end{multline*}
where $u_1,u_2,u_3,u_4$
are four coefficients 
that can take any value in $\{0,1\}$.

According to \Tref{thm:subset}, the subgroup $\LTA(5,2)_f$ acts transitively 
on $\cX$. Since the group action of the affine transformations 
in $\LTA(5,2)_f$ can be viewed as permutations 
on the codeword coordinates, we can conclude that 
all 16 cosets in $\cX$ have the same weight distribution.
\label{ex:G32}

\vspace{1mm}
\noindent \textbf{\textit{Computing the Weight Distribution}}
\vspace{1mm}
	
If we directly apply Algorithm \ref{alg:WEF_polar_dyn} 
to compute the weight distribution of $\mbC$, 
we need to compute the weight enumerating function 
of $2^9$ polar cosets. Now we show how we can reduce this number 
using \Tref{thm:subset}.

We start by partitioning code $\mbC$ into two parts according to $u_6$. 
Let $\mbC\{u_6 = 1\}$ denote the subset of $\mbC$ where $u_6$ is fixed to be 1, 
and let $\mbC\{u_6 = 0\}$ denote the subcode of $\mbC$ where $u_6$ is fixed to be 0. 
Then $$\mbC = \mbC\{u_6 = 1\} \cup \mbC\{u_6 = 0\},$$
and both $\mbC\{u_6 = 1\}$ and $\mbC\{u_6 = 0\}$ 
can be represented as disjoint unions of $2^8$ polar cosets.
Next, we compute the weight distribution of 
$\mbC\{u_6 = 1\}$ and $\mbC\{u_6 = 0\}$ separately.

For $\mbC\{u_6 = 1\}$, observe that 
$$\mbC\{u_6 = 1\} = \bigcup_{X\in\cX} X$$
By \Tref{thm:subset}, all cosets in $\cX$ have the same weight distribution. 
Thus to get the weight distribution for $\mbC\{u_6 = 1\}$, 
it suffices to first compute the weight distribution of a single coset in 
$\cX$ using Algorithm \ref{alg:WEF_polar_dyn}, and then multiply it by $|\cX| = 2^4$.
This reduces the number of polar cosets that we need to evaluate 
for $\mbC\{u_6 = 1\}$ from $2^8$ down to $2^4$.

After that we consider $\mbC\{u_6 = 0\}$. 
Since the subcode $\mbC\{u_6 = 0\}$ is also a decreasing 
monomial code itself, we can again partition $\mbC\{u_6 = 0\}$ 
into two parts according to $u_7$:
$$\mbC\{u_6 = 0\} = 
\mbC\{u_6 = 0, u_7 = 1\} \cup
\mbC\{u_6 = 0, u_7 = 0\},$$
and apply \Tref{thm:subset} to reduce the complexity 
for the second term $\mbC\{u_6 = 0, u_7 = 1\}$. 

By repeating this procedure, code $\mbC$ can be unfolded as follows
\begin{align*}
\mbC =\; &\mbC\{u_6 = 1\} \\
\cup\; &\mbC\{u_6 = 0, u_7 = 1\} \\
\cup\; &\mbC\{u_6 = 0, u_7 = 0, u_9 = 1\} \\
\cup\; &\cdots \\
\cup\; &\mbC\{u_6 = 0, u_7 = 0, \cdots, u_{14} = 0, u_{15} = 1\} \\
\cup\; &\mbC\{u_6 = 0, u_7 = 0, \cdots, u_{14} = 0, u_{15} = 0\}, 
\end{align*}
and \Tref{thm:subset} allows us to reduce the number of polar cosets that 
we need to evaluate for each of those components. 
The amount of complexity reduction for the components, 
and the total amount of complexity reduction for $\mbC$ 
are shown in Table \ref{tb:ex7_1}. 
In Table \ref{tb:ex7_1}, the second column shows the number of polar cosets 
in each component, and the third column shows the number of polar cosets 
that we need to evaluate after applying \Tref{thm:subset}.
We also show the computed weight distribution 
for $\mbC$ in this example in Table \ref{tb:ex7_2},
where the unlisted $A_d$ equals to zero.
\begin{table}[!t]
\renewcommand{\arraystretch}{1.3}
\caption{The complexity reduction amounts in Example \ref{ex:G32}}
\label{tb:ex7_1}
\centering
\begin{tabular}{|l|l|l|}
\hline
components & complexity &  reduced complexity \\
\hline
$\mbC\{u_6 = 1\}$               & $2^8 = 256$ & $2^4 = 16$ \\ \hline
$\mbC\{\cdots, u_7 = 1\}$       & $2^7 = 128$ & $2^3 = 8$\\ \hline
$\mbC\{\cdots, u_9 = 1\}$       & $2^6 = 64$ &  $2^2 = 4$\\ \hline
$\mbC\{\cdots, u_{10} = 1\}$    & $2^5 = 32$ &  $2^2 = 4$\\ \hline
$\mbC\{\cdots, u_{11} = 1\}$    & $2^4 = 16$ &  $2^1 = 2$\\ \hline
$\mbC\{\cdots, u_{12} = 1\}$    & $2^3 = 8$ &   $2^1 = 2$\\ \hline
$\mbC\{\cdots, u_{13} = 1\}$    & $2^2 = 4$ &   $2^0 = 1$\\ \hline
$\mbC\{\cdots, u_{14} = 1\}$    & $2^1 = 2$ &   $2^0 = 1$\\ \hline
$\mbC\{\cdots, u_{15} = 1\}$    & $2^0 = 1$ &   $2^0 = 1$\\ \hline
$\mbC\{\cdots, u_{15} = 0\}$    & $2^0 = 1$ &   $2^0 = 1$\\ \hline
\hline
$\mbC$    & $2^9=512$ & 40\\ \hline
\end{tabular}
\end{table}
\begin{table}[!t]
\renewcommand{\arraystretch}{1.3}
\caption{Weight distribution of $\mbC$ in Example \ref{ex:G32}}
\label{tb:ex7_2}
\centering
\begin{tabular}{|r|l|r|l|r|l|}
\hline
$d$ & $A_d$  & $d$ & $A_d$    & $d$ & $A_d$   \\\hline
0   & 1      & 12  & 1768424  & 22  & 503424  \\\hline
4   & 472    & 14  & 3668224  & 24  & 83164   \\\hline
6   & 6272   & 16  & 4717254  & 26  & 6272    \\\hline
8   & 83164  & 18  & 3668224  & 28  & 472     \\\hline
10  & 503424 & 20  & 1768424  & 32  & 1       \\\hline
\end{tabular}
\end{table}
\end{example}

\subsection{Proof of \Tref{thm:subset}}
The rest of this section is devoted to the proof of \Tref{thm:subset}. 
First let us introduce more notations. For any polynomial $p\in\cR_m$, 
we can expand it and express $p$ as a sum of monomials in $\cM_m$ as
\begin{equation}
p = \sum_{q\in \cM_m} u_q \cdot q, 
\label{eq:p_expansion}
\end{equation}
where $u_q\in\{0,1\}$ are the coefficients. 
For each monomial $q$, we denote the coefficient $u_q$ in this 
expansion of $p$ by $\langle p \rangle_q$. Using this notation, 
equation (\ref{eq:p_expansion}) can be written as
$$p = \sum_{q\in \cM_m} \langle p\rangle_q \cdot q,$$
We start our proof by establishing a few lemmas. 
First, we consider the group action of an affine transformation $(A,\bfb)$ 
in the subgroup $\LTA(m,2)_f$ on $f$ itself. 
The following lemma states that the coefficient of a monomial 
$h\in\cS$ in the expansion of $(A,\bfb)\cdot f$ 
can actually be determined by an entry in $(A,\bfb)$.
\begin{lemma}
In \Tref{thm:subset}, let $(A,\bfb)\in\LTA(m,2)_f$ with $A = (a_{ij})$, 
and $h\in\cS$, then the coefficient of $h$ in the expansion of $(A,\bfb)\cdot f$ 
equals to an entry in $(A,\bfb)$. More precisely, 
\begin{itemize}
    \item if $f = qx_s$ and $h = qx_t$ for some monomial $q$ with 
    $t < s$, then $\langle (A,\bfb)\cdot f\rangle_h = a_{st}$;
    \item if $f = hx_s$, 
    then $\langle (A,\bfb)\cdot f\rangle_h = b_{s}$. 
\end{itemize}
\label{lm:A_b_f}
\end{lemma}
\begin{IEEEproof}
First, $f$ can be written as
$$f = \prod_{i\in\ind(f)}x_i$$
Consider the action of $(A,\bfb)\in\LTA(m,2)$ on $f$. 
According to Definition \ref{def:LTA_f}, 
each monomial $x_i$ in $f$ will be replaced by 
$$y_i = x_i + \sum_{j<i:\; j\not\in\ind(f)}a_{ij}x_j + b_i$$
Therefore, 
$(A,\bfb)\cdot f$ can be written as a product of $\ell$ linear terms, 
where $\ell = \deg f$:
\begin{equation}
(A,\bfb)\cdot f = \prod_{i\in\ind(f)}
\left(
x_i + \sum_{j<i:\; j\not\in\ind(f)}a_{ij}x_j + b_i
\right)
\label{eq:l_linear_forms_v1}
\end{equation}
Given that $h\in\cS$ is a one-variable descendant of $f$, 
we now verify this lemma by discussing 
the following two cases for $h$:
\begin{itemize}
\item Case 1: $f = qx_s$ and $h = qx_t$ for some monomial $q$ with $t < s$. 

It can be observed that when we expand the right hand side of equation (\ref{eq:l_linear_forms_v1}), 
there is only one way to generate the term $h$, 
corresponding to choosing $a_{st}x_t$ from the linear term led by $x_s$, 
and choosing the leading $x_i$ for the rest of the linear terms. 
Thus $\langle (A,\bfb)\cdot f \rangle_h = a_{st}$. 

\item Case 2: $f = hx_s$. 

Similarly, it can be observed that when we expand the 
right hand side of equation (\ref{eq:l_linear_forms_v1}), 
there is only one way to generate the term $h$, 
corresponding to choosing $b_{s}$ from the linear term  
led by $x_s$, and choosing the leading $x_i$ for 
the rest of the linear terms. Thus $\langle (A,\bfb)\cdot f \rangle_h = b_{s}$. 
\end{itemize}
\end{IEEEproof}

Next, we consider the group action of an $(A,\bfb)\in\LTA(m,2)_f$ on a monomial $g\in\cT$. 
The following lemma states that, the coefficient of a monomial $h\in\cS$ 
in the expansion of $(A,\bfb)\cdot g$ always equals to zero.
\begin{lemma}
In \Tref{thm:subset}, let $(A,\bfb)\in\LTA(m,2)_f$, $h\in\cS$, and $g\in\cT$, 
then the coefficient of $h$ in the expansion of $(A,\bfb)\cdot g$ is zero. 
In other words, $\langle (A,\bfb)\cdot g\rangle_h = 0$.
\label{lm:coeff}
\end{lemma}
\begin{IEEEproof}
Consider the action of $(A,\bfb)$ on $g$. According to Definition \ref{def:LTA_f}, 
the monomials in $g$ will change as follows:
\begin{itemize}
\item Every $x_i$ with $i\in\ind(g)\cap\ind(f)$ will be replaced by
$$y_i = x_i + \sum_{j<i,\;j\not\in\ind(f)}a_{ij}x_j + b_i$$
\item Every $x_i$ with $i\in\ind(g)\backslash\ind(f)$ will 
be replaced by $y_i = x_i$, and thus remain unchanged.
\end{itemize}
So after the action by $(A,\bfb)$ on 
$$
g = \prod_{i\in\ind(g)}x_i,
$$
we have
\begin{multline}
   (A,\bfb)\cdot g = 
   \underbrace{
   \left(
   \prod_{i\in\ind(g)\backslash\ind(f)}x_i
   \right)
   }_{(a)} \\
   \cdot\underbrace{
   \left(
   \prod_{i\in\ind(g)\cap\ind(f)}
   \left(
   x_i+\sum_{j<i,\;j\notin\ind(f)}a_{ij}x_j
   +b_i
   \right)
   \right)
   }_{(b)}
   \label{eq:lm_f_g}
\end{multline}
If $\langle (A,\bfb)\cdot g\rangle_h = 1$, then $h$ should appear 
if we expand the right hand side of (\ref{eq:lm_f_g}). 
Since $h\in\cS$ is a one-variable descendant of $f$, 
according to Definition \ref{def:ovd}, we have the following 
two possible cases for the relation between $h$ and $f$. 
Next, we show that if $\langle (A,\bfb)\cdot g\rangle_h = 1$, 
a contradiction can be drawn in both cases.
\begin{itemize}
\item Case 1: $f = qx_s$ and $h = qx_t$ for some monomial $q$ with $t<s$. 

If $h$ appears in the expansion of the right hand side of (\ref{eq:lm_f_g}), 
then we break it into two cases depending on where the $x_t$ in $h$ comes from. 
\begin{itemize}
\item 
If the $x_t$ in $h$ comes from parenthesis (a) in (\ref{eq:lm_f_g}), 
then we must have $\ind(g)\backslash\ind(f) = t$,
and $\ind(q) \subseteq \ind(g)\cap\ind(f)$.
Since $q$ is a divisor of $f$, for $\ind(q) \subseteq \ind(g)\cap\ind(f)$ to be true, 
we can either have $$\ind(q) = \ind(g)\cap\ind(f) \quad\Rightarrow\quad g = h$$
which is a contradiction since $g$ and $h$ are distinct, or
$$\ind(q)\cup\{s\} = \ind(g)\cap\ind(f)
\quad\Rightarrow\quad g = qx_tx_s,$$
which is also a contradiction since $\rowind{g} > \rowind{f}$.
\item 
If the $x_t$ in $h$ comes from parenthesis (b) in (\ref{eq:lm_f_g}), 
then we must have $\ind(g)\backslash\ind(f) = \emptyset$, and 
$\ind(q)\cup\{s\} \subseteq \ind(g)\cap\ind(f)$, 
giving us $$\ind(g) = \ind(f) \quad\Rightarrow\quad g = f,$$
which is a contradiction since $g$ and $f$ are distinct.
\end{itemize}
\item 
Case 2: $f = hx_s$.

If $h$ appears in the expansion of the right hand side of (\ref{eq:lm_f_g}), 
then we must have $\ind(g)\backslash\ind(f) = \emptyset$, 
and $\ind(h) \subseteq \ind(g)\cap\ind(f)$.
Since $h$ is a divisor of $f$, for $\ind(h) \subseteq \ind(g)\cap\ind(f)$ to be true, 
we can either have
$$\ind(h) = \ind(g)\cap\ind(f) \quad\Rightarrow\quad g = h,$$
which is a contradiction since $g$ and $h$ are distinct,
or
$$\ind(h)\cup\{s\} = \ind(g)\cap\ind(f) \quad\Rightarrow\quad g = f$$
which is also a contradiction since $g$ and $f$ are distinct.
\end{itemize}
Therefore, in both Case 1 and Case 2, $\langle (A,\bfb)\cdot g\rangle_h = 1$ 
leads to a contradiction.
Thus we can only have $\langle (A,\bfb)\cdot g\rangle_h = 0$.
\end{IEEEproof}

Using \Lref{lm:coeff}, we can prove that subcode $\mbC(\cT)$ 
is invariant under $\LTA(m,2)_f$, as stated in the following lemma.
\begin{lemma}
In \Tref{thm:subset}, subcode $\mbC(\cT)$ is invariant under $\LTA(m,2)_f$.
\label{lm:CT_invariant}
\end{lemma}
\begin{IEEEproof}
Let $(A,\bfb)\in\LTA(m,2)_f$. The group action by $(A,\bfb)$ 
can be viewed as a permutation on the codeword coordinates, 
so $(A,\bfb)$ acting on $\mbC(\cT)$ will generate another subspace with 
the same dimension as $\mbC(\cT)$.
Since $\mbC(\cT)$ is generated by the monomials in $\cT$, 
to prove this claim, it suffices to prove that for any $g\in\cT$, 
we have $(A,\bfb)\cdot\ev(g)\in\mbC(\cT)$.

Let $(A,\bfb)\in\LTA(m,2)_f$ and $g\in\cT$. 
First, it follows from \Pref{pp:Abf} that
\begin{equation}
(A,\bfb)\cdot g = g + \sum_{g'\in\cM_m:\;g'\prec g} u_g' \cdot g',
\label{eq:Abg_1}
\end{equation}
where $u_g'\in\{0,1\}$ are coefficients for all $g'$.
Then, since $\cI$ is the generating set of a 
decreasing monomial code, from \Tref{BDOT-degradation} we know all $g'$ with 
$g'\prec g$ are in $\cI$. Hence (\ref{eq:Abg_1}) can be written as 
\begin{equation}
(A,\bfb)\cdot g = g + \sum_{g'\in\cI:\;g'\prec g} u_g' \cdot g'. 
\label{eq:Abg_2}
\end{equation}

Recall $f$ is the monomial with the smallest row index in $\cI$, 
so it follows from \Lref{lm:total_order_is_a_linear_extension} that
$f\not\preceq g$. Also, \Lref{lm:coeff} tells us that in (\ref{eq:Abg_2}), 
$u_h = 0$ for all $h\in\cS$. Since $\cI = \{f\}\cup\cS\cup\cT$, 
(\ref{eq:Abg_2}) becomes
\begin{equation*}
(A,\bfb)\cdot g = g + \sum_{g'\in\cT:\;g'\prec g} u_g' \cdot g'
\end{equation*}
Therefore, any $(A,\bfb)\cdot g$ with 
$g\in\cT$ can be generated by 
the monomials in $\cT$. 
This finishes the proof of this lemma.
\end{IEEEproof}

At this point, we are ready to put everything together and 
prove \Tref{thm:subset}. Take  $X_0 = \ev(f) + \mbC(\cT)$
to be a coset in $\cX$. To prove that the group action of $\LTA(m,2)_f$ 
on $\cX$ is transitive, it suffices to prove that 
the orbit of $X_0$ is the entire $\cX$.

Let $(A,\bfb)$ be an affine transformation in $\LTA(m,2)_f$. 
If we consider the action of $(A,\bfb)$ on $f$, 
It follows from \Pref{pp:Abf} that
\begin{equation}
(A,\bfb)\cdot f = f + \sum_{h\in\cS} u_h\cdot h + 
\sum_{g\in\cT} u_g\cdot g
\label{eq:f_uh_ug}
\end{equation}
where $u_h = \langle (A,\bfb)\cdot f \rangle_h$ for each $h\in\cS$, 
and $u_g = \langle (A,\bfb)\cdot f \rangle_g$ for each $g\in\cT$.
Therefore, if we look at the action of $(A,\bfb)$ on $X_0$, 
we have
\begin{align}
&(A,\bfb)\cdot X_0 \nonumber\\
&= 
\ev(f) + \sum_{h\in\cS} u_h\cdot \ev(h) + 
\sum_{g\in\cT} u_g\cdot \ev(g) + (A,\bfb)\cdot\mbC(\cT) 
\label{eq:thm6_1}\\
&= 
\ev(f) + \sum_{h\in\cS} u_h\cdot \ev(h) + 
\sum_{g\in\cT} u_g\cdot \ev(g) + \mbC(\cT) 
\label{eq:thm6_2}\\
&= 
\ev(f) + \sum_{h\in\cS} u_h\cdot \ev(h) + \mbC(\cT) \\
&= 
\ev(f) + \sum_{h\in\cS} \langle (A,\bfb)\cdot f \rangle_h\cdot \ev(h) + \mbC(\cT)
\label{eq:thm6_3}
\end{align}
where 
\begin{itemize}
    \item 
    in (\ref{eq:thm6_1}), we have $(A,\bfb)\cdot\mbC(\cT) = \mbC(\cT)$, 
    since $(A,\bfb)\in\LTA(m,2)_f$, and $\mbC(\cT)$ is invariant 
    under $\LTA(m,2)_f$ from \Lref{lm:CT_invariant}.
    \item
    in (\ref{eq:thm6_2}), we have
    $$
    \sum_{g\in\cT} u_g\cdot \ev(g) + \mbC(\cT) = \mbC(\cT),
    $$
    since $\ev(g) \in \mbC(\cT)$ for all $g\in\cT$.
\end{itemize}

In (\ref{eq:thm6_3}), according to \Lref{lm:A_b_f}, 
every $\langle (A,\bfb)\cdot f\rangle_h$ equals to an 
entry in $(A,\bfb)$. Therefore, given any $X\in\cX$, 
we can pick an $(A,\bfb)\in\LTA(m,2)_f$ whose entries 
are chosen such that $X$ can be generated by $(A,\bfb)\cdot X_0$.
This proves that the orbit of $X_0$ is the entire set $\cX$, 
which means the group action of $\LTA(m,2)_f$ on the $\cX$ is transitive.
Since the action by affine transformations in $\LTA(m,2)_f$ 
can be viewed as permutations on the codeword coordinates, 
all the cosets in $\cX$ thus have the same weight distribution.
This completes the proof.

\section{Applications of Our Algorithm}
In this section, we present some application examples 
of our algorithm on decreasing monomial codes. 
For those codes, we also show how much the complexity 
of our approach can be reduced using the method from Section~\ref{sec:LTA}.

First, we present the weight distribution of a $(128,64)$ polar code 
constructed based on the reliability sequence in the 5G 
technical specification \cite{3GPP} without CRC. 
This polar code has mixing factor 34, 
so if we directly apply Algorithm \ref{alg:WEF_polar_dyn}, 
the number of polar cosets that we need to evaluate equals $2^{34}$.
The code can be verified to be a 
decreasing monomial code, so this complexity can be reduced
using the method from Section~\ref{sec:LTA}. 
After the complexity reduction, the number of polar cosets that 
we need to evaluate can be reduced to $39257360 \approx 2^{25.23}$. 
The computed weight distribution of this code is shown in 
Table \ref{tb:N128_K64_5G_WD}. 
For reference, computing this weight distribution 
takes less than two hours on a laptop computer.
\begin{table}[!t]
\renewcommand{\arraystretch}{1.3}
\caption{Weight distribution of the (128,64) polar code 
constructed following the reliability sequence in 5G \cite{3GPP}}
\label{tb:N128_K64_5G_WD}
\centering
\scalebox{0.9}{
\begin{tabular}{|r|l|}
\hline
$d$  & $A_d$\\\hline
  0  & 1\\
  8  & 304\\
 12  & 768\\
 16  & 161528\\
 20  & 4452096\\
 24  & 166137744\\
 28  & 8299319808\\
 32  & 474588991516\\
 36  & 19910428320256\\
 40  & 555627871531568\\
 44  & 9459383897458944\\
 48  & 94101946507153608\\
 52  & 550051775557674240\\
 56  & 1920378732932218128\\
 60  & 4051638142931561472\\
 64  & 5194332067339587654\\
 68  & 4051638142931561472\\
 72  & 1920378732932218128\\
 76  & 550051775557674240\\
 80  & 94101946507153608\\
 84  & 9459383897458944\\
 88  & 555627871531568\\
 92  & 19910428320256\\
 96  & 474588991516\\
100  & 8299319808\\
104  & 166137744\\
108  & 4452096\\
112  & 161528\\
116  & 768\\
120  & 304\\
128  & 1\\\hline
\end{tabular}
}
\end{table}

Then, we look at the (128,64) Reed-Muller code. Note that 
the weight distribution of this Reed-Muller code has already been 
computed by Sugino, Ienaga, Tokura and Kasami in \cite{sugino1971weight}.
This Reed-Muller code has mixing factor 49, so in our approach, 
the number of polar cosets that we need to evaluate in 
Algorithm \ref{alg:WEF_polar_dyn} equals $2^{49}$. 
If we apply the complexity reduction from Section \ref{sec:LTA},
this number can be reduced to $49761365064 \approx 2^{35.53}$, 
which is a viable computation complexity that can be achieved. 
Since this self-dual Reed-Muller code has the largest mixing factor among all 
decreasing monomial codes with rate at most 1/2 at length 128. 
It is reasonable to expect that after we apply the complexity reduction 
from Section \ref{sec:LTA}, the number of polar cosets that
we need to evaluate for other decreasing monomial codes
at length 128 will not be much larger than $2^{35}$.
Therefore, we believe that our approach allows us to compute 
the weight distribution of any decreasing monomial codes at length 128.

\section{Conclusion}
In this paper, we present a deterministic algorithm for computing the exact 
weight distribution of polar codes at length 128. 
First, we propose a recursive procedure for computing the 
weight distribution of polar cosets along arbitrary 
decoding path. Then, we show that any polar code can be represented as 
a disjoint union of polar cosets.
Therefore, the entire weight distribution of the code 
can be obtained by first computing the weight distribution of 
all the polar cosets in this representation, and then taking the sum.
However, the number of polar cosets in this representation grows 
exponentially with a parameter called mixing factor.
To bound the complexity of our approach, we provide 
a bound on the mixing factor of polar codes 
being decreasing monomial codes. To further reduce this complexity, 
we study the algebraic structure of decreasing monomial codes, 
and prove that a subgroup of the lower triangular affine group 
acts transitively on certain subsets of 
decreasing monomial codes. This allows us to reduce the 
number of polar cosets that we need to evaluate in our approach.
After the complexity reduction, our algorithm still has exponential complexity, 
but it is efficient enough to compute the weight distribution 
of any decreasing monomial codes at length 128.



%

\begin{IEEEbiographynophoto}{Hanwen Yao}
(Member, IEEE) received his B.S. degree in electrical engineering from 
the Shanghai Jiao Tong University in Shanghai, China, in 2016, 
and his M.S. and Ph.D. degrees in electrical engineering 
from the University of California, San Diego (UCSD) in 2018 and 2022, respectively. 
He received the Jack Keil Wolf ISIT Student Paper Award in 2021. 
He was also awarded the Shannon Graduate Fellowship 
by the Center for Memory and Recording Research in UCSD in 2021. 
He is currently a Postdoctoral Associate 
in the Department of Electrical and Computer Engineering with Duke University. 
His research interests include coding theory, combinatorics, and algorithm. 
\end{IEEEbiographynophoto}


\begin{IEEEbiographynophoto}{Arman Fazeli}
(Member, IEEE) received the B.S. degree in electrical engineering 
from the Sharif University of Technology, Tehran, Iran, in 2012, and 
the Ph.D. degree in electrical and computer engineering from the University of
California at San Diego. He was a Post-Doctoral Scholar and a Lecturer at the
University of California at San Diego until 2021. He is currently a Wireless
System Engineer at Apple. His main research interests include coding and
information theory, with emphasis on polar codes and coding for distributed
storage systems. He received Silver and Bronze Medals at the International
Mathematical Olympiad (IMO) in 2006 and 2007, when he was in the Iran
National Mathematics Team.
\end{IEEEbiographynophoto}

\begin{IEEEbiographynophoto}{Alexander Vardy}
(Fellow, IEEE) was born in Moscow, Russia, in 1963.
He received the B.Sc. degree ({\it summa cum laude}) from the Technion—Israel
Institute of Technology, Israel, in 1985, and the Ph.D. degree from Tel-Aviv
University, Israel, in 1991.

From 1985 to 1990, he was with the Israeli Air Force, where he worked
on electronic counter measures systems and algorithms. From 1992 to 1993,
he was a Visiting Scientist with the IBM Almaden Research Center, San
Jose, CA, USA. From 1993 to 1998, he was with the University of Illinois at
Urbana-Champaign, first as an Assistant Professor and then as an Associate
Professor. Since 1998, he has been with the University of California at San
Diego (UCSD), where he is currently the Jack Keil Wolf Chair Professor
with the Department of Electrical and Computer Engineering and the Department of Computer Science. While on sabbatical from UCSD, he has held
long-term visiting appointments with CNRS, France, the EPFL, Switzerland,
the Technion—Israel Institute of Technology, and Nanyang Technological
University, Singapore. His research interests include error-correcting codes,
algebraic and iterative decoding algorithms, lattices and sphere packings,
coding for storage systems, cryptography, computational complexity theory,
and fun math problems. He was a member of the Board of Governors of the
IEEE Information Theory Society from 1998 to 2006 and from 2011 to 2017.
In 1996, he became a fellow of the David and Lucile Packard Foundation.
He received the IBM Invention Achievement Award in 1993 and the NSF
Research Initiation and CAREER Awards in 1994 and 1995. In 1996, he was
appointed as a fellow of the Center for Advanced Study, University of
Illinois, and received the Xerox Award for Faculty Research. He received the
IEEE Information Theory Society Paper Award (jointly with Ralf Koetter)
in 2004. In 2005, he received the Fulbright Senior Scholar Fellowship and
the Best Paper Award at the IEEE Symposium on Foundations of Computer
Science (FOCS). In 2017, his work on polar codes was recognized by
the IEEE Communications and Information Theory Societies Joint Paper
Award. From 1995 to 1998, he was an Associate Editor for {\it Coding Theory}.
From 1998 to 2001, he was the Editor-in-Chief of the 
{\sc IEEE Transactions on Information Theory}.
\end{IEEEbiographynophoto}




\end{document}